\documentclass[11pt,a4paper]{article}
\usepackage{epsfig,graphicx,amsfonts,amsmath,amssymb}

\usepackage{cite}
\RequirePackage{mathrsfs}

\newlength{\dinwidth}
\newlength{\dinmargin}
\def\eq#1{{Eq.~(\ref{#1})}}
\newcommand{\Le}{\left(}
\newcommand{\Ra}{\right)}

\newcommand{\beq}{\begin{equation}}
\newcommand{\eeq}{\end{equation}}
\newcommand{\beqar}{\begin{eqnarray}}
\newcommand{\eeqar}{\end{eqnarray}}
\newcommand{\D}{\partial}

\newcommand{\ep}{\varepsilon}

\newcommand{\tv}{\textsl{v}}

\newcommand{\T}{{\cal T}}
 
\date{}
\begin{document}

\title {{~}\\
{\Large \bf  On reggeization of vertex of three reggeized gluons in high energy QCD }}
\author{ 
{~}\\
{\large 
S.~Bondarenko$^{(1) }$,
S.~Pozdnyakov$^{(1) }$
}\\[7mm]
{\it\normalsize  $^{(1) }$ Physics Department, Ariel University, Ariel 40700, Israel}\\
}

\maketitle
\thispagestyle{empty}

\begin{abstract}

 The  unitarity corrections to the propagator of reggeized gluons calculated in the framework of QCD RFT require a knowledge of the
expressions for  Reggeon propagator  and vertices of interaction of three reggeized gluons (Reggeons) to one QCD loop precision, see \cite{Our6}. In this paper we calculate the vertex of 
interactions of $A_+ A_+ A_-$
Reggeon fields, i.e. vertex of transition of $A_-$ Reggeon field to  $A_+ A_+$ fields, to this precision. We demonstrate, that all loop leading logarithmic order contributions to the vertex
can be summed through the integro-differential equation similarly to the BFKL one,  \cite{BFKL}. The solution of this equation leads to the reggeized from of the vertex 
with the trajectory twice larger then the trajectory of reggeized gluon propagator. The application of the obtained result is also discussed.
  
\end{abstract}

\section{Introduction}

 $\,\,\,\,\,\,$ The  high energy, effective QCD action for an interaction of the reggeized gluons (Reggeons), introduced in \cite{LipatovEff,LipatovEff1},  see also \cite{Our1,Our2,Our3,Our4,Our41,Our5},
describes quasi-elastic amplitudes of high-energy scattering processes
in the  multi-Regge kinematics. The applications of the approach to the description of  high energy processes
and calculation of unitarity  corrections to the different production 
amplitudes can be found in \cite{EffAct,Fadin,Nefedov} for example, whereas
the generalization of the formalism for a case of the description of arbitrary production amplitudes and impact factors is presented in \cite{Our3} with
the prescription of the calculation of $S$-matrix elements accordingly to an approach of \cite{Faddeev}.
This effective action formalism, based on the reggeized gluons as main degrees of freedom, see \cite{BFKL}, can be considered as reformulation of the RFT (Regge Field Theory) calculus introduced
in \cite{Gribov}, see also \cite{0dim1,0dim2,0dim3,0dim4,Pom1,Pom2,Pom22,Pom3,Pom4}, for the case of high energy QCD.
It was underlined in \cite{LipatovEff,LipatovEff1} that the main purposes of the approach is the construction of the
$S$-matrix unitarity in the direct and crossing channels of the scattering processes through the multi-Reggeon dynamics described by the  vertices  of multi-Reggeon
interactions, see simirar approaches in \cite{Venug,Kovner,Hatta1,BKP,GLR,BK,TripleV}, the connections between the different formalisms were clarified in \cite{Our1,Our5,Hetch}. 
The unitarity of the Lipatov's formalism, therefore, is related to 
the unitarity corrections in both RFT and QCD sectors of the theory.

 Similarly to the phenomenological theories of interacting Reggeons, see \cite{Gribov,0dim1,0dim2,0dim3,0dim4} and references therein, we can separately calculate the corrections 
to the amplitudes which come from the pure RFT sector of the formalism. Namely, let us consider  Lipatov's effective action
for reggeized gluons $A_{\pm}$ formulated as  RFT (Regge Field Theory) in the form of generatin functional obtained by an integration out of the gluon fields $v$ from the
the $S_{eff}[v,\,A]$:
\beq\label{Add1}
e^{\imath\,\Gamma[A]}\,=\,\int\,D v\,e^{\imath\,S_{eff}[v,\,A\,]}
\eeq
where
\beq\label{Add2}
S_{eff}\,=\,-\,\int\,d^{4}\,x\,\Le\,\frac{1}{4}\,G_{\mu \nu}^{a}\,G^{\mu \nu}_{a}\, \,+\,tr\,\left[\,\Le\,\T_{+}(v_{+})\,-\,A_{+}\,\Ra\,j_{reg}^{+}\,+\,
\Le\,\T_{-}(\textsl{v}_{-})\,-\,A_{-}\,\Ra\,j_{reg}^{-}\,\right]\,\Ra\,,
\eeq
with
\beq\label{Add3}
\T_{\pm}(v_{\pm})\,=\,\frac{1}{g}\,\D_{\pm}\,O(v_{\pm})\,=\,
v_{\pm}\,O(v_{\pm})\,,\,\,\,\,
j_{reg\,a}^{\pm}\,=\,\frac{1}{C(R)}\,\D_{i}^{2}\,A_{a}^{\pm}\,,
\eeq
here  $C(R)$ is eigenvalue of Casimir operator in the representation R, $tr(T^{a} T^{b})\,=\,C(R)\,\delta^{a b}$\, see \cite{LipatovEff,LipatovEff1,Our1,Our2}.
The form of the Lipatov's operator  $O$ (and correspondingly $\T$) depends on the particular process of interests, see \cite{Our4},
we take it in the form of  the Wilson line (ordered
exponential) for the longitudinal gluon fields in the adjoint representation:
\beq\label{Add4}
O(v_{\pm})\,=\,P\,e^{g\,\int_{-\infty}^{x^{\pm}}\,d x^{\pm}\,v_{\pm}(x^{+},\,x^{-},\,x_{\bot})}\,,\,\,\,\,\tv_{\pm}\,=\,\imath\,T^{a}\,v_{\pm}^{a}\,,
\eeq
see also \cite{Nefedov}. 
There are additional kinematical constraints for the reggeon fields
\beq\label{Ef4}
\partial_{-}\,A_{+}\,=\,\partial_{+}\,A_{-}\,=\,0\,,
\eeq
corresponding to the strong-ordering of the Sudakov components in the multi-Regge kinematics,
see \cite{LipatovEff,LipatovEff1,Our4}. The action is constructed by the request that the  LO value of the classical gluon fields in the solutions of equations of motion will be fixed as
\beq\label{Ef8}
\textsl{v}_{\pm}^{cl}\,=\,A_{\pm}\,.
\eeq
In the light-cone gauge $\textsl{v}_{-}\,=\,0$,  the equations of motion can be solved and the general expressions for the gluon fields can be written in the following form:
\beq\label{Ef9}
v_{i}^{a}\,\rightarrow\,v^{a}_{i\,cl}(A_{\pm})\,+\,\ep_{i}^{a}\,,\,\,\,\,v_{+}^{a}\,\rightarrow\,v_{+\,cl}^{a}(A_{\pm})\,+\,\ep_{+}^{a}\,.
\eeq
The integration in respect to the fluctuations around the classical solutions provides QCD loop corrections to the effective vertices of the Lipatov's action which now can be written as functional of 
the Reggeon fields only\footnote{In order to make the notations shorter, we change the position of the color and other indexes of the vertices further in the article, preserving only the
overall number of the indexes.} :
\beq\label{Sec1}
\Gamma = \sum_{n,m\,=\,1}\Le\,A_{+}^{\,a_1}\cdots A_{+}^{\,a_n} \Le K^{+\,\cdots\,+}_{-\,\cdots\,-}\Ra^{a_1\cdots\, a_n}_{b_1\cdots\, b_m} A_{-}^{\,b_1} \cdots\,A_{-}^{\,b_m} \Ra = 
-\,A_{+\,x}^{\,a} \D_{i}^{2}\,A_{-\,x}^{\,a} + A_{+\,x}^{\,a} \Le K_{x y}^{\,a\, b} \Ra^{+}_{-} A_{-\,y}^{\,b} + \cdots\,,
\eeq
in general  the summation 
on the color indexes in the r.h.s of the equation means the integration on the corresponding coordinates as well.
Now we see, that the theory have two different sources of any perturbative/unitarity corrections. The first one comes from the QCD sector of the formalism, it affects on the precision
of the effective vertices (kernels)
calculated in the pure QCD. Another source of the corrections is described by the processes formulated in terms of RFT sector degrees of freedom only, i.e. these corrections
are constructed entirely in terms of the Reggeon fields and 
\eq{Sec1} vertices known to some QCD precision. These type of the  RFT corrections were considered many times
in the previous phenomenological RFT approaches, see \cite{0dim1,0dim2,0dim3,0dim4} and references therein for example; in the effective high energy QCD formalism there are the RFT corrections to the propagator and vertices of the \eq{Sec1} action as well. In the paper \cite{Our5} the Dyson-Schwinger hierarchy of the equations for the correlators of reggeized gluons was derived in the 
framework of the formalism 
that allows to determine the calculation scheme for these corrections to any correlator of interests. We also note, that the obtained hierarchy  formally is similar to the Balitsky hierarchy
of equations and BK-JIMWLK approaches , see \cite{Venug,Kovner,Hatta1,BK}, and there is a correspondence between different degrees of freedom such as  reggeized gluons and Wilson line operators, 
see details in \cite{Our5,Hetch}.

 In our previous paper, \cite{Our6}, we first time calculated the one RFT loop correction to the propagator of reggeized gluons. It turns out, that this correction is large and 
does not suppressed perturbatively. Namely, beginning from the NLO there are terms of the same order in the correction which are involved in the full answer with the sign opposite to 
the leading order term sign, see details in \cite{Our6}. The answer obtained is an example of the non-linear correction to the propagator of reggeized gluons, similar in some extend to the
non-linear corrections of \cite{BK} formalism. The non-linear correction calculated in \cite{Our6}, nevertheless, was obtained with the use of the bare triple Reggeon vertices only. In order to 
provide the full one loop QCD precision for the RFT correction, we have to know the one QCD loop expression for the triple Reggeon vertices as well\footnote{See \cite{White}
where similar vertex was considered.}. 
Therefore, in this paper we calculate the one QCD loop correction to the $A_{+} A_{+} A_{-}$ vertex of interaction of three reggeized gluons\footnote{The calculation of the $A_{-} A_{-} A_{+}$ vertex is in the progress.}. It is found, that we can write for the vertex the equation of the Bethe - Salpeter type which sum up all loopscontributions to the vertex to the LLA precision. The solution of this equation leads to the reggeization of the vertex with the trajectory twice large than the trajectory of the propagator of reggeized gluons, this is a main result of the article. 
Consequently, the paper is organized as follows. In the next section we remind some basic definitions from the \cite{Our2}, this Section as well as Appendices A-C 
intended to facilitate the understanding of the technical calculations. The Section 3 is dedicated to the calculation of the one loop QCD
correction to the vertex of interests, the main technical details of the calculations are located in Appendix D. The last Section is the Conclusion of the article
where we discuss the result of the calculations and their possible applications.

\section{One loop effective action}

The general expressions for the gluon fields can be written in the following form:
\beq\label{Ef1}
v_{i}^{a}\,\rightarrow\,v^{a}_{i\,cl}\,+\,\ep_{i}^{a}\,,\,\,\,\,v_{+}^{a}\,\rightarrow\,v_{+\,cl}^{a}\,+\,\ep_{+}^{a}\,,
\eeq
at the next step we expand the Lagrangian of the effective action around this classical solution. Preserving in the expression only terms which are quadratic with respect to the fluctuation fields,  
we obtain for this part of the action:
\beqar\label{Ef2}
S_{\ep^{2}}& = &-\frac{1}{2}\,\int\,d^{4} x\,\left.\Big( \ep_{i}^{a}\Le \delta_{ac}\Le \delta_{ij}\,\Box\, +\D_{i}\,\D_{j} \Ra - \right.\right. \nonumber \\
&-&
2 g f_{abc} \Le \delta_{ij} \,v_{k}^{b\,cl} \D_{k}  -
2\, v_{j}^{b\,cl} \D_{i}  + v_{i}^{b\,cl} \D_{j} - \delta_{ij} \,v_{+}^{b\,cl} \D_{-}  \Ra    -\nonumber\\
&-&\left.\,
\,g^{2}\,f_{a b c_{1}}\,f_{c_{1} b_{1} c}\,
\Le \delta_{i j}\,v_{k}^{b\,cl}\,v_{k}^{b_{1}\,cl}\,
-\,v_{i}^{b_1\,cl}\,v_{j}^{b\,cl}\,\Ra\,\Ra\,\ep_{j}^{c}\,+\,\nonumber\\
&+&\,
\,\ep_{+}^{a}\Le -2 \,\delta^{a c}\,\D_{-}\D_{i} -2 g f_{abc} \Le  v_{i}^{b\,cl} \D_{-} - \Le \D_{-} v_{i}^{b\,cl}\Ra \Ra \Ra\,\ep_{i}^{c}\,
+\,\nonumber\\
&+&\,\left.\,
\ep_{+}^{a}\, \delta_{a c}\, \D_{-}^{2} \,\ep_{+}^{c} -
\,g\,\ep_{+\,x}^{a}\,\int\,d^{4} y\,\Le U^{a\,b\,c}_{1} \Ra^{+}_{x\,y} \Le \D_{i} \D_{-} \rho_{b}^{i}\Ra_{x}\,\ep_{+\,y}^{c}\,\Ra
\,=\,\nonumber \\
&=&\,-\,\frac{1}{2}\,\ep_{\mu}^{a}\,\Le\,\Le M_{0} \Ra_{\mu\, \nu}^{ac}\,+\,\Le M_{1} \Ra_{\mu\, \nu}^{ac}\,+\,
\Le M_{2} \Ra_{\mu\, \nu}^{ac}\,+\,\Le M_{L} \Ra_{\mu\, \nu}^{ac}\Ra\,\ep_{\nu}^{c}\,.
\eeqar
Here we defined $\Le M_{i} \Ra_{\mu\, \nu}^{ac}\,\propto\,g^{i}$ and note that
\beq\label{Ef3}
\Le M_{1}\Ra_{-\,i}\,=\,-\, g f_{abc} \Le  v_{i}^{b\,cl} \,\overrightarrow{\D_{-}} - \Le \D_{-} v_{i}^{b\,cl}\Ra \Ra\,,\,\,\,
\Le M_{1}\Ra_{i\,-}\,=\,-\, g f_{abc} \Le  \,\overleftarrow{\D_{-}}\,  v_{i}^{b\,cl}- \Le \D_{-} v_{i}^{b\,cl}\Ra \Ra\,.
\eeq
The last term in \eq{Ef2} expression, denoted as $\Le M_{L} \Ra_{\mu\, \nu}^{ac}$ represents contribution of the Lipatov's effective current into the action. 
This term is defined trough the following function:
\beq\label{Ef411}
\Le U_{1}^{a\, b\, c} \Ra_{x y}^{+}\,=\,
tr[\,f_{a}\,G^{+}_{x y}\,f_{c}\,O_{y}\,f_{b}\,O^{T}_{x}]\,+\,
tr[\,f_{c}\,G^{+}_{y x}\,f_{a}\,O_{x}\,f_{b}\,O^{T}_{y}]\,,
\eeq
see definitions of the quantities in Appendix B. We underline, that \eq{Ef4} function can be  expanded as an infinite series in respect to $g$ coupling constant and $v_{+}^{cl}$ fields,
see again Appendix B. Now we can perform the integration obtaining the one loop effective action:
\beqar\label{Ef5}
\Gamma\,& = &\,\int\,d^{4} x\,\Le L_{YM}(v_{i}^{cl},\, v_{+}^{cl})- v_{+\,cl}^{a}\,J_{a}^{+}(v_{+}^{cl})- A_{+}^{a}\,\Le\,\D_{i}^{2}\,A_{-}^{a}\,\Ra\,\Ra + \nonumber \\
& + &\,
\frac{\imath}{2}\,Tr\,\ln\Le\,\delta_{\rho\,\nu}\,+\,G_{0\,\rho\,\mu}\,\Le\, \Le M_{1}\Ra_{\mu\,\nu}\,+\,\Le M_{2}\Ra_{\mu\,\nu}\,+\,
\Le M_{L}\Ra_{\mu\,\nu}\,\Ra\,\Ra\, + \nonumber \\
&+&\,\frac{1}{2}\,\int\,d^{4} x\,\int\,d^{4} y\,
j_{\mu\,x}^{\,a}\,G_{\mu\,\nu}^{\,a b}(x,y)\,j_{\nu\,y}^{\,b}\,.
\eeqar
Here we have: $\,G_{\,0\,\nu\,\mu}$ as bare  gluon propagator
\beq\label{Ef6}
\Le M_0\Ra^{\,\mu\,\nu}\,G_{\,0\,\nu\,\rho}\,=\,\delta_{\mu \rho}\,,
\eeq
see Appendix A; the full gluon propagator is defined as
\beq\label{Ef7}
G_{\mu \nu}^{ac}\,=\,\left[\,\Le M_{0} \Ra_{\mu \nu}^{ac}\,+\,\Le M_{1} \Ra_{\mu \nu}^{ac}\,+\,\Le M_{2} \Ra_{\mu \nu}^{ac}\,+\,\Le M_{L}\Ra_{\mu\,\nu}^{ac}\,\right]^{-1}\,
\eeq
and can be written in the form of the following  perturbative series:
\beq\label{Eff8}
G_{\mu \nu}^{ac}(x,y)\,=\,G_{0\,\mu \nu}^{ac}(x,y)\,-\,\int\,d^4 z\,G_{0\,\mu \rho}^{ab}(x,z)\,
\Le \,\Le M_{1}(z)\Ra_{\rho \gamma}^{bd}\,+\,\Le M_{2}(z)\Ra_{\rho \gamma}^{bd}\,+\,\Le M_{L}(z)\Ra_{\rho\,\gamma}^{bd}\, \Ra G_{\gamma \mu}^{dc}(z,y)\,;
\eeq
the auxiliary currents $j_{\mu\,x}^{\,a}$ and  $j_{\nu\,y}^{\,b}$ are requested for the many-loops calculations of the effective action, in our case 
of calculation of one loop precision we take them equal zero from the beginning.

\section{Reggeization of $K^{++-}$ triple Reggeon vertex}

 Any  vertex of the three Reggeon fields interactions  is defined as following:
\beq\label{BV1}
\Le K^{a b c }_{x y z}\Ra^{\mu \nu \rho}\,=\,\int\, d^{4} w\, \Le\,\frac{\delta^{3}\,\Gamma(v_{i}^{cl}(A),\,v_{+}^{cl}(A))}{\delta A_{\mu}^{a}(x)\,\delta A_{\nu}^{b}(y)\, \delta A_{\rho}^{c}(z) }\,\Ra_{A\,=\,0}\,,\,\,\,\,\,\,\mu \nu \rho=(+,\,-)\,
\eeq
The bare triple Reggeon vertices were calculated in \cite{Our6}, the one loop corrections to the  $K^{++-}$ vertex are presented in Appendix D. Summing up \eq{D21}, \eq{D36} and \eq{D43}
expressions we obtain for the vertex to one QCD loop precision:
\beqar\label{OL1}
&\,&-2\,\imath\, \Le K^{a\,b\,c}_{x\,z\,y\,}\Ra^{+  + -}_{1}\,=\, \\
&=&\,-\,
\frac{\imath\,g^3\,N}{2 (2 \pi)^{5}}\,f_{a b c}\,\delta^{2}_{x_{\bot}\,z_{\bot}}\,
\D_{i\,y}^{2}\,\Le\,
\theta(z^{+} - x^{+})\,\int \frac{d k_{-}}{k_{-}}\,
\int d^2 k_{\bot} \,
\int d^{2} k_{1\,\bot}\,\frac{k_{1\,\bot}^{2}}{k_{\bot}^2\,\Le k_{\bot} - k_{1\,\bot} \Ra^2}\,\,e^{-\imath\,(y^{i}\,-\,z^{i})\,k_{1\,i}}\,-\, \right. \nonumber \\
&-&\,\left.
\theta(x^{+} - z^{+})\,\int \frac{d k_{-}}{k_{-}}\,\int d^{2} k_{\bot}\,
\int d^2 k_{\bot} \,
\int d^{2} k_{1\,\bot}\,\frac{k_{1\,\bot}^{2}}{k_{\bot}^2\,\Le k_{\bot} - k_{1\,\bot} \Ra^2}\,e^{-\imath\,(y^{i}\,-\,x^{i})\,k_{1\,i}}\,\,
\Ra\, \nonumber .
\eeqar  
Correspondingly, performing \eq{D1314} variable's change,  we write the vertex in the following form:
\beqar\label{OL3}
&\,& \Le K^{a\,b\,c}_{x\,z\,y}\Ra^{+ + -}_{1}=\frac{g^3\,N}{2\,(2 \pi)^{5}}\,f_{a b c}\,\delta^{2}_{x_{\bot}\,z_{\bot}}\,\eta\,\D_{i\,y}^{2}\, 
\Le\,\theta(z^{+}\,-\,x^{+})\, \int d^2 k_{\bot} \, 
\int d^{2} k_{1\,\bot}\,\frac{k_{1\,\bot}^{2}}{k_{\bot}^2\,\Le k_{\bot} - k_{1\,\bot} \Ra^2}\,\,e^{-\imath\,(y^{i}\,-\,z^{i})\,k_{1\,i}}\,-\,
\right. \nonumber \\
&-&\left.
\theta(x^{+}\,-\,z^{+})\, 
\int d^2 k_{\bot} \,
\int d^{2} k_{1\,\bot}\,\frac{k_{1\,\bot}^{2}}{k_{\bot}^2\,\Le k_{\bot} - k_{1\,\bot} \Ra^2}\,e^{-\imath\,(y^{i}\,-\,x^{i})\,k_{1\,i}}\,
\Ra\,.
\eeqar
Using it's bare value 
\beq\label{OL4}
\Le K^{a b c }_{x z y}\Ra_{0}^{+ + -}\,=\,\frac{1}{2}\,g\,f^{a b c }\,
\Le \theta (x^{+}-y^{+})\,-\,\theta (z^{+}-x^{+}) \Ra\, \delta^{2}(y_{\bot}\,-\,x_{\bot})\,\delta^{2}(z_{\bot}\,-\,y_{\bot})\,\D_{i\,y}^{2}\,,
\eeq
see \cite{Our6} for the details of the calculation, we can write the following integral equation for the vertex to the LLA precision:
\beqar\label{OL31}
&\,& 
\Le K^{a\,b\,c}_{x\,z\,y;\,\eta}\Ra^{+ + -}\,=\,\Le K^{a b c }_{x z y}\Ra_{0}^{+ + -}\,-\, \\
&-&\,
\frac{2\,\alpha_s\,N}{(2\pi)^4}\,\int_{0}^{\eta}\, ds\,\int\,d^2 w_{\bot}\,\Le K^{a\,b\,c}_{x\,z\,w;\,s}\Ra^{+ + -}\,
\int d^2 k_{\bot} \,
\int d^{2} k_{1\,\bot}\,\frac{k_{1\,\bot}^{2}}{k_{\bot}^2\,\Le k_{\bot} - k_{1\,\bot} \Ra^2}\,e^{-\imath\,(y^{i}\,-\,w^{i})\,k_{1\,i}}\,.
\eeqar
The leading logarithmic approximation (LLA) structure of the equation will be clear if we put attention that 
\beq\label{OL32}
\D_{i y}^{4}\,\int \,dt\,D_{+ -\, 0}(w_{\bot}, t_{\bot}) D_{+ -\, 0}(t_{\bot}, y_{\bot}) D_{+ -\, 0}(t_{\bot}, w_{\bot})= 
\int \frac{d^2 k_{\bot}}{(2 \pi)^2} 
\int \frac{d^{2}  k_{1\,\bot}}{(2 \pi)^2} \frac{k_{1\,\bot}^{2}}{k_{\bot}^2\,\Le k_{\bot} - k_{1\,\bot} \Ra^2} e^{-\imath\,(y^{i}\,-\,w^{i})\,k_{1\,i}}\, \nonumber 
\eeq
is an one loop correction to the bare vertex, 
see Appendix C for the definition of $D_{+ -\, 0}$.
Now, performing Fourier transform with respect to the transverse variables of the full vertex
\beq\label{OL5}
\Le K^{a\,b\,c}_{x\,z\,y;\,\eta}\Ra^{+ + -}\,=\,\int\, \frac{d^2 p}{(2\pi)^2}\,\int\, \frac{d^2 p_1}{(2\pi)^2}\,\int\, \frac{d^2 p_2}{(2\pi)^2}\,
\tilde{K}^{a\,b\,c}(x^+ , z^+ ; p_{\bot},p_{1\bot} ,p_{2 \bot}; \eta)\,e^{-\imath\,p_{i}x^i\,-\,\imath\, p_{1\,i} z^i\,-\,p_{2\,i} y^{i}}\,
\eeq
and taking derivative on $\eta$ we obtain:
\beq\label{OL511}
\frac{\D \tilde{K}^{a\,b\,c}}{\D \eta}\,=\,2\,\varepsilon (p_{2 \bot})\,\tilde{K}^{a\,b\,c}\,
\eeq
with
\beq
\varepsilon (p_{\bot})\,=\,-\,\frac{\alpha_s\,N}{4\,\pi^2}\,\int d^{\,2} k_{\bot} \,\frac{p_{\bot}^{2}}{k_{\bot}^2\,\Le k_{\bot} - p_{\bot} \Ra^2}\,.
\eeq\label{OL6}
The solution of the \eq{OL5} with given precision is the following function:
\beq\label{OL7}
\tilde{K}^{a\,b\,c}(x^+ , z^+ ; p_{\bot},p_{1\bot} ,p_{2 \bot}; \eta)\,=\,\tilde{K}^{a\,b\,c}_{0}(x^+ , z^+ ; p_{\bot},p_{1\bot} ,p_{2 \bot}; \eta)\,
e^{2\,\eta\,\varepsilon (p_{2\,\bot})}\,,
\eeq
which can be considered as reggeization of the bare vertex \eq{OL4}, see \eq{Pro11} expression as well.

\section{Conclusion}

 In the formalism of Lipatov's effective action, formulated as RFT,  there is an additional source of perturbative and unitarity corrections to high energy QCD amplitudes based
on the diagrams constructed entirely in terms of Reggeon fields and their vertices of interactions. The completeness of the correction to the propagator of regggeized gluon,
calculated in \cite{Our6}, requires the knowledge of the vertices of the interaction of three Reggeon fields to one QCD loop precision. In the paper we calculated the $A_{+} A_{+} A_{-}$
vertex with all loop LLA precision, see \eq{OL7} expression which is the main result of the article.

 The bare QCD value of the vertex of interests also was calculated in \cite{Our6}. Writing precise one QCD loop expression for the vertex, see \eq{OL3}, we can put attention that the
many loops contribution to the vertex are reproduced by iterations expressed finally in the form of Bethe-Salpeter equation. The solution of this equation, \eq{OL7}, have a form of reggeized vertex 
with the trajectory  twice larger than the trajectory of the Reggeon's propagator, this is a new and unexpected result of the calculations. The high energy QCD reformulated as RFT, therefore, 
becomes non-local in rapidity space with both vertices and propagators as functions of rapidity intervals. Of course, this is result of the addition of the Lipatov's effectice currents to the 
pure QCD Lagrangian, whereas these currents are absent these non-local terms disappear as well. Also, we note, that the calculated terms describe the high-energy asymptotic behavior of the theory,
there are additional QCD type contributions to any functions of interests which provide sub-leading corrections to both vertices and propagators.   

 We also note, that in general the Dyson-Schwinger  hierarchy for the vertices of the formalism exists as well, similarly to the hierarchy of the theory's correlators obtained in \cite{Our5}. 
The derivation of this sytem of equations, as well as calculation of the $A_{-} A_{-} A_{+}$ vertex to one QCD loop precision will allow to determine the next leading order non-linear corrections to the propagator of reggeized gluon, which is important task in high energy QCD. Also,
the next important step to be considered is the calculation of the BFKL Pomeron on the base of new expression for the reggeized gluons propagator, see \cite{Our5}. Indeed, the infrared divergence of 
obtained in \cite{Our6} propagator is different from the divergence of the usual propagator's trajectory function, the situation will be even worse when the both triple vertices will be 
included in the answer with one QCD loop precision. Additionally, an interesting question arises about the possible reggeization of the four Reggeon vertex in the framework of the theory. 
Therefore, the interesting subjects of the 
future research are the non-linear corrections to the Reggeons  propagator and Pomeron calculated in the framework with possible reggeization of the vertices of the theory included.
For example,  the very interesting question to investigate is about the form and rapidity dependence of this modified Pomeron.

 In conclusion we emphasize, that  the article is considered as an additional step to the developing of the high energy QCD RFT which will help clarify the non-linear unitarity corrections to the 
amplitudes of high energy processes.

\newpage
\section*{Appendix A: Bare gluon propagator in light-cone gauge}
\renewcommand{\theequation}{A.\arabic{equation}}
\setcounter{equation}{0}

 In order to find the expression for the gluon fields bare propagator\footnote{We suppress color and coordinate notations in the definition of the propagators below.} in the light-cone gauge we solve the following system of equations
\beq\label{A1}
M^{\,0\,\mu\,\nu}\,G_{\,0\,\nu\,\rho}\,=\,\delta^{\,\mu}_{\,\rho}
\eeq
with
\beq\label{A101}
g^{\,\mu}_{\,\nu}\, =\,\left(
\begin{array}{cccc}
0 & 1 & 0 & 0 \\
1 & 0 & 0 & 0 \\
0 & 0 & 1 & 0 \\
0 & 0 & 0 & 1
\end{array} \right)\,\,\,\,\,\mu\,,\nu\,=\,(+,\,-,\,\bot)\,,
\eeq
see also \eq{A13} below.
The expression for the $M_{\,0\,\mu\,\nu}$ matrix can obtained from the bare gluon's Lagrangian for the gluon's fluctuations field, in light-cone gauge it has the following form:
\beq\label{A11}
L_{0}\, = \,-\,\frac{1}{2}\,\ep_{i}^{a}\,\delta_{\,a\,b}\Le \delta_{ij}\,\Box\, +\D_{i}\,\D_{j} \Ra\,\ep_{j}^{b}\,+\,
\ep_{+}^{a}\,\D_{-}\,\D_{i}\,\ep_{i}^{a}\,-\,\frac{1}{2}\,\ep_{+}^{a}\,\D_{-}^{2}\,\ep_{+}^{a}\,=\,
-\,\frac{1}{2}\,\ep_{\mu}^{a}\,M_{\,0\,\,\mu\,\nu}\,\ep_{\nu}^{b}\,\delta^{\,a\,b}\,,
\eeq
In the following system of equations 
\beqar\label{A12}	
\,M^{\,i\,+}_{\,0}\,G_{\,0\,+\,j}\,+\,M^{\,i\,k}_{\,0}\,G_{\,0\,k\,j}\,=\,\delta^{\,i}_{\,j}\,\nonumber \\
\,M^{\,+\,i}_{\,0}\,G_{\,0\,i\,+}+M^{\,+\,+}_{\,0}\,G_{\,0\,+\,+}=\delta^{\,+}_{\,+}\,\nonumber \\
\,M^{\,+\,i}_{\,0}\,G_{\,0\,i\,j}\,+\,M^{\,+\,+}_{\,0}\,G_{\,0\,+\,j}\,=\,0\, \nonumber \\
\,M^{\,i\,p}_{\,0}\,G_{\,0\,p\,+}\,+\,M^{\,i\,+}_{\,0}\,G_{\,0\,+\,+}\,=\,0\,,
\eeqar
the last two equations we can consider as definitions of corresponding Green's functions:
\beq\label{A2}	
G_{0\,+\,i}\,=\,-M_{0\,+\,+}^{-1}\,\,M^{\,+\,j}_{\,0}\,G_{\,0\,j\,i}\,,
\eeq	
and
\beq\label{A21}
G_{0\,i\,+}\,=\,-M_{0\,i\,j}^{-1}\,\,M^{\,j\,+}_{\,0}\,G_{\,0\,+\,+}\,.
\eeq	
Here for
\beq\label{A3}
M_{\,0\,\,p\,j}\,=\,\delta_{p\,j}\,\Box\, +\D_{p}\,\D_{j}\,,\,\,\,M_{\,0\,p\,-}\,=\,-\,\D_{p}\,\D_{-}\,,\,\,\,M_{\,0\,\,-\,-}\,=\,\D_{-}^{2}\,
\eeq
we have
\beq\label{A211}
M^{-1}_{0\,ij}(x,y)\,=\,-\,\int\,\frac{d^4 p}{(2\pi)^{4}}\,\frac{e^{-\imath\,p\,(x\,-\,y)}}{p^2}\,\Le
\delta_{ij}\,-\,\frac{p_{i}\,p_{j}}{2\Le p_{-}\,p_{+}\Ra}\Ra\,
\eeq
and correspondingly
\beq\label{A22}
M^{-1}_{0\,+\,+}(x,y)\,=\,-\,\int\,\frac{d^4 p}{(2\pi)^{4}}\,\frac{e^{-\imath\,p\,(x\,-\,y)}}{p^{2}_{-}}\,.
\eeq
Therefore, for the two remaining Green's functions we obtain:
\beq\label{A2211}
\Le\, M^{+\,+}_{\,0}\,-\,M_{\,0}^{\,+\,i}\,M_{0\,i\,j}^{-1}\,\,M^{\,j\,+}_{\,0}\,\Ra\,G_{\,0\,+\,+}\,=\,\delta^{\,+}_{\,+}\,,
\eeq
and
\beq\label{A23}
\Le\, M^{i\,k}_{\,0}\,-\,M_{\,0}^{\,i\,+}\,M_{0\,+\,+}^{-1}\,\,M^{\,+\,k}_{\,0}\,\Ra\,G_{\,0\,k\,j}\,=\,\delta^{\,i}_{\,j}\,.
\eeq
Performing Fourier transform of the functions, we write the \eq{A23} in the following form:
\beq\label{A24}
-\int\,\frac{d^4 p}{(2\pi)^{4}}\,\Le \delta^{\,i\,k}\,p^{\,2} + p^{\,i}\,p^{\,k}  \Ra\,e^{-\imath\,p\,(x\,-\,y)}\,\tilde{G}_{\,0\,k\,j}(p)\,+\,
\int\,\frac{d^4 p}{(2\pi)^{4}}\,p_{-}^{\,2}\,p^{\,i}\,p^{\,k}\,\frac{e^{-\imath\,p\,(x\,-\,y)}}{p_{-}^{\,2}}\,\tilde{G}_{\,0\,k\,j}(p)\,=\,\delta^{\,i}_{\,j}\,
\eeq
that provides:
\beq\label{A25}
G_{0\,i\,j}(x,y)\,=\,-\,\int\,\frac{d^4 p}{(2\pi)^{4}}\,\frac{e^{-\imath\,p\,(x\,-\,y)}}{p^{\,2}}\,\delta_{\,i\,j}\,.
\eeq
Correspondingly, for \eq{A22} we have:
\beq\label{A26}
-\int\,\frac{d^4 p}{(2\pi)^{4}}\,p_{\,-}^{\,2}\,e^{-\imath\,p\,(x\,-\,y)}\,\tilde{G}_{\,0\,+\,+}(p)\,+\,
\int\,\frac{d^4 p}{(2\pi)^{4}}\,\frac{p_{\,-}^{\,2}\,p^{\,i}\,p^{\,j}}{p^{\,2}} \Le \delta_{\,i\,j}\,-\,\frac{p_{i}\,p_{j}}{2\Le p_{-}\,p_{+}\Ra} \Ra\,e^{-\imath\,p\,(x\,-\,y)}\,
\tilde{G}_{\,0\,+\,+}(p)\,\,=\,\delta^{\,+}_{\,+}\,,
\eeq
that can be rewritten as
\beq\label{A261}
\int\,\frac{d^4 p}{(2\pi)^{4}}\,e^{-\imath\,p\,(x\,-\,y)}\,\Le
-\,p_{\,-}^{\,2}\,+\,p_{\,-}^{\,2}\,\frac{p^{\,i}\,p^{\,i}}{p^{\,2}}\,-\,p_{\,-}^{\,2}\,\frac{\Le p^{\,i}\,p_{\,i}\Ra\,\Le p^{\,j}\,p_{\,j}\Ra}{2\,p^{\,2} \Le p_{\,+} p_{\,-} \Ra}\,\Ra\,
\tilde{G}_{\,0\,+\,+}(p)\,\,=\,\delta^{\,+}_{\,+}\,.
\eeq
Writing this expression as
\beq\label{A262}
\int\,\frac{d^4 p}{(2\pi)^{4}}\,e^{-\imath\,p\,(x\,-\,y)}\,\Le
-\,p_{\,-}^{\,2}\,-\,p_{\,-}^{\,2}\,\frac{p^{\,i}\,p_{\,i}}{p^{\,2}}\,-\,p_{\,-}^{\,2}\,\frac{\Le p^{\,i}\,p_{\,i}\Ra\,\Le p^{\,j}\,p_{\,j}\Ra}{2\,p^{\,2} \Le p_{\,+} p_{\,-} \Ra}\,\Ra\,
\tilde{G}_{\,0\,+\,+}(p)\,\,=\,\delta^{\,+}_{\,+}\,.
\eeq
we obtain finally for the Green's function
\beq\label{A263}
G_{0\,+\,+}(x,y)\,=\,-\,\int\,\frac{d^4 p}{(2\pi)^{4}}\,\frac{e^{-\imath\,p\,(x\,-\,y)}}{p^{\,2}}\,\frac{2\,p_{\,+}}{p_{\,-}}\,.
\eeq
Inserting \eq{A25} and \eq{A263} functions in \eq{A2}-\eq{A21} definitions we obtain for the last two Green's functions:
\beq\label{A5}
G_{\,0\,i\,+}\,=\,G_{\,0\,+\,i}\,=\,\int\,\frac{d^4 p}{(2\,\pi)^{4}}\,\frac{e^{-\,\imath\,p\,(x\,-\,y)}}{p^{\,2}}\,\frac{p_{\,i}}{p_{-}}\,.
\eeq	
Now, introducing the following vector in light-cone coordinates
\beq\label{A1111}
n_{\,\mu}^{+}\,=\,(1,\,0,\,0_{\bot})\,,\,\,\,\,\mu\,=\,(+,\,-,\,\bot)
\eeq
 we can write the whole propagator as
\beq\label{A1211}
G_{\,0\,\mu\,\nu}(x,\,y)\,=\,\int\,\frac{d^{4} p}{(2\,\pi)^{4}}\,\frac{e^{-\,\imath\,p\,(x\,-\,y)}}{p^{\,2}}\,\Le\,g_{\mu\,\nu}\,-\,g_{\mu\,\sigma}\,g_{\nu\,\rho}\,
\frac{p^{\,\sigma}\,n^{\,+\,\rho}\,+\,p^{\,\rho}\,n^{\,+\,\sigma}\,}
{p^{\,\rho}\,n_{\,\rho}^{\,+}}\,\Ra\,
\eeq                         
with
\beq\label{A13}
g_{\,\mu\,\nu}\, =\, g^{\,\mu\,\nu}\,= \left(
\begin{array}{cccc}
0 & 1 & 0 & 0 \\
1 & 0 & 0 & 0 \\
0 & 0 & -1 & 0 \\
0 & 0 & 0 & -1
\end{array} \right)\,\,\,\,\,\mu\,,\nu\,=\,(+,\,-,\,\bot)\,
\eeq
and where Kogut-Soper convention, \cite{}, for the light-cone notations and scalar product is used:
\beq\label{A31}
p\,x\,=\,p_{+}\,x^{+}\,+\,p_{-}\,x^{-}\,+\,p_{i}\,x^{i}\,.
\eeq

\newpage
\section*{Appendix B: Lipatov's effective current}
\renewcommand{\theequation}{B.\arabic{equation}}
\setcounter{equation}{0}

For the arbitrary representation of gauge field $v_{+}\,=\,\imath\,T^{a}\,v_{+}^{a}$ with
$D_{+}\,=\,\D_{+}\,-\,g\,v_{+}$, we can consider
the following representation of $O$ and $O^{T}$ operators
\footnote{Due the light cone gauge we consider here only $O(x^{+})$ operators. 
The construction of the representation of the $O(x^{-})$ operators can be done similarly. We also note, that the integration is assumed for 
repeating indexes in expressions below if it is not noted otherwise. }:
\beq\label{B11}
O_{x}\,=\,\delta^{a\, b}\,+\,g\,\int\,d^{4}y\,G_{x y}^{+\,a\, a_{1}}\, \Le v_{+}(y)\Ra_{a_1\,b} \, =\,
\,1\,+\,g\,G_{x y}^{+}\, v_{+ y}\,
\eeq
and correspondingly
\beq\label{B12}
O^{T}_{x}\,=\,\,1\,+\,g\,v_{+ y}\,G_{y x}^{+}\,,
\eeq
which is redefinition of the operator expansions used in \cite{LipatovEff} in terms of Green's function instead 
integral operators, see Appendex B above.
The Green's function in above equations we understand as Green's function of the $D_{+}$ operator
and express it in the perturbative sense as :
\beq\label{B13}
G_{x y}^{+}\,=\,G_{x y}^{+\,0}\,+\,g\,G_{x z}^{+\,0}\, v_{+ z}\,G_{z y}^{+}\,
\eeq
and
\beq\label{B14}
G_{y x}^{+}\,=\,G_{y x}^{+\,0}\,+\,g\,G_{y z}^{+}\, v_{+ z}\,G_{z x}^{+\,0}\,,
\eeq
with the bare propagators defined as (there is no integration on index $x$ in expressions)
\beq\label{B15}
\D_{+ x}\,\,G_{x y}^{+\,0}\,=\,\delta_{x\,y}\,,\,\,\,G_{y x}^{+\,0}\,\overleftarrow{\D}_{+ x}\,=\,-\delta_{x\,y}\,.
\eeq
The following properties of the operators now can be derived:
\begin{enumerate}
\item
\beqar\label{B161}
\delta\,G^{+}_{x y}& = & g\,G_{x z}^{+\,0}\,\Le \delta v_{+ z} \Ra\,G_{z y}^{+}+
\,G_{x z}^{+\,0}\, v_{+ z}\,\delta G_{z y}^{+}=
g\,G_{x z}^{+\,0}\,\Le \delta v_{+ z} \Ra\,G_{z y}^{+}+
\,G_{x z}^{+\,0}\, v_{+ z}\,\Le \delta G_{z p}^{+} \Ra\,D_{+ p}\,G^{+}_{p y}=\nonumber \\
&=&
g \Le
G_{x z}^{+\,0}\,\Le \delta v_{+ z} \Ra\,G_{z y}^{+}-
G_{x z}^{+\,0}\, v_{+ z}\,G_{z p}^{+}\,\Le \delta D_{+ p}\Ra\,G^{+}_{p y}\Ra=
g \Le
G_{x p}^{+\,0}\,+\,G_{x z}^{+\,0}\, v_{+ z}\,G_{z p}^{+}\Ra\,\delta v_{+ p} \,G^{+}_{p y}=\nonumber \\
&=&\,
\,g\,G^{+}_{x p}\,\delta v_{+ p} \,\,G^{+}_{p y}\,;
\eeqar
\item
\beq\label{B16}
\delta\,O_{x}\, = \,g\,G^{+}_{x y}\,\Le \delta v_{+ y} \Ra\,+g\,\Le \delta G^{+}_{x y}\Ra\,v_{+ y}\,=
\,g\,G^{+}_{x p}\,\delta v_{+ p}\,\Le 1\, +\,g \,G^{+}_{p y}\,v_{+ y}\,\Ra\,=\,
g\,G^{+}_{x p}\,\delta v_{+ p}\,O_{p}\,;
\eeq
\item
\beq\label{B17}
\D_{+ x}\,\delta\,O_{x}\,=\,g\,\Le \D_{+ x}\,G^{+}_{x p} \Ra\,\delta v_{+ p}\,O_{p}\,=\,
g\,\Le 1\,+\,g\,\,v_{+ x}\,G^{+}_{x p}\,\Ra\,\delta v_{+ p}\,O_{p}\,=\,
g\,O^{T}_{x}\,\delta v_{+ x}\,O_{x}\,;
\eeq
\item
\beq\label{B18}
\D_{+ x}\,O_{x}\,=\,g\,\Le \D_{+ x}\,G^{+}_{x y} \Ra\,v_{+ y}\,=\,
g\,v_{+ x}\,\Le 1\,+\,g\,G_{x y}^{+}\,v_{+ y}\,\Ra\,=\,
g\,v_{+ x}\,O_{x}\,;
\eeq
\item
\beq\label{B19}
O_{x}^{T}\,\overleftarrow{\D}_{+ x}\,=\,g\,v_{+ y}\,\Le G^{+}_{y x}\,\overleftarrow{\D}_{+ x} \Ra\,=\,
-\,g\,\Le 1\,+\,v_{+ y}\,\,G^{+}_{y x}\,\Ra\,v_{+ x}\,=\,-g\,O^{T}_{x}\,v_{+ x}\,.
\eeq
\end{enumerate}
We see, that the operator $O$ and $O^{T}$ have the properties of ordered exponents. For example, choosing bare propagators as
\beq\label{B110}
\,G_{x y}^{+\,0}\,=\,\theta(x^{+}\,-\,y^{+})\,\delta^{3}_{x y}\,,\,\,\,
\,G_{y x}^{+\,0}\,=\,\theta(y^{+}\,-\,x^{+})\,\delta^{3}_{x y}\,,\,\,\,
\eeq
we immediately reproduce:
\beq\label{B111}
O_{x}\,=\,P\, e^{g\int_{-\infty}^{x^{+}}\,dx^{'+}\, v_{+}(x^{'+})} \,,\,\,\,
O^{T}_{x}\,=\,P\, e^{g\int_{x^{+}}^{\infty}\,dx^{'+}\, v_{+}(x^{'+})} \,.
\eeq
The form of the bare propagator
$
\,G_{x y}^{+\,0}\,=\,\frac{1}{2}\,\left[\,\theta(x^{+}\,-\,y^{+})\,-\,\theta(y^{+}\,-\,x^{+})\,\right]\,\delta^{3}_{x y}\,
$ 
will lead to the
more complicated representations of $O$ and $O^{T}$ operators, see in \cite{LipatovEff} and \cite{Our4}.
We note also that the Green's function notation $\tilde{G}^{\pm\,0}_{x^{\pm} y^{\pm}}$ in the paper is used for the designation of the only theta function part of the full $G^{\pm 0}_{x^{\pm} y^{\pm}}$ Green's function.

 Now we consider a variation of the action's full current :
\beq\label{B112}
\delta\, tr[v_{+ x}\,O_{x}\,\D_{i}^{2}\,A^{+}]=
\frac{1}{g}\,\delta\, tr[\Le \D_{+ x}\,O_{x} \Ra \D_{i}^{2}\,A^{+}]=\frac{1}{g}\, tr[
\Le\D_{+ x} \delta \,O_{x} \Ra \D_{i}^{2}\,A^{+}] = tr[
O^{T}_{x}\,\delta v_{+ x}\,O_{x}\Le \D_{i}^{2}\,A^{+}\Ra]\,,
\eeq 
which can be rewritten in the familiar form used in the paper:
\beq\label{B1121}
\delta\,\Le v_{+}\,J^{+} \Ra\,=\delta\,tr[\, \Le\, v_{+ x}\,O_{x}\,\D_{i}^{2}\,A^{+}\,\Ra\,]\,=\,-\,
\delta v_{+}^{a}\,tr[\,T_{a}\,O\,T_{b}\,O^{T}\,]\,\Le \D_{i}^{2} A^{+}_{b}\Ra\,.
\eeq
We also note, that with the help of \eq{A11} representation of the $O$ operator
the full action's  current can we written as follows
\beq\label{B113}
 tr[\Le v_{+ x}\,O_{x}\,-\,A_{+}\Ra\,\D_{i}^{2}\,A^{+}\,]\,=\,
tr[\Le v_{+}\,-\,A_{+} + v_{+ x}\,G^{+}_{x y}\,v_{+ y}\,\Ra\,\Le \D_{i}^{2} A^{+}\Ra]\,.
\eeq

\newpage
\section*{Appendix C: NLO vertex of interactions of reggeized gluons}
\renewcommand{\theequation}{C.\arabic{equation}}
\setcounter{equation}{0}

 The NLO one-loop vertex of reggeized gluons interactions is defined in the formalism as
\beqar\label{C1}
&-&2\,\imath\, K^{a\,b}_{x\,y\,1}\,=\,\Le\,\frac{\delta^{2}\,\ln\Le 1 + G_0\,M\,\Ra}{\delta A_{+\,x}^{a}\,\delta A_{-\,y}^{b}}\,\Ra_{A_{+},\,A_{-},\,v_{f\,\bot}\,=\,0}\,=\,\nonumber \\
&=&\left[ G_0\,\frac{\delta^{2}\,M}{\delta A_{+\,x}^{a} \delta A_{-\,y}^{b}} \Le 1 + G_0\,M\,\Ra^{-1}-
 G_0\,\frac{\delta\,M}{\delta A_{-\,y}^{b}}\Le 1 + G_0\,M\Ra^{-1}
 G_0\,\frac{\delta\,M}{\delta A_{+\,x}^{a}}\Le 1 + G_0\,M\Ra^{-1}
\right]_{A_{+},A_{-},v_{f\,\bot} =\, 0 }\,
\eeqar
where the trace of the expression is assumed. With the help of \eq{Ef5}, see also \cite{Our2}, we have correspondingly:
\beq\label{C2}
-2\,\imath\, K^{a\,b}_{x\,y\,1}\,=\,
\left[ G_0\,\frac{\delta^{2}\,M}{\delta A_{+\,x}^{a} \delta A_{-\,y}^{b}} -
 G_0\,\frac{\delta\,M}{\delta A_{-\,y}^{b}}
 G_0\,\frac{\delta\,M}{\delta A_{+\,x}^{a}}
\right]_{A_{+},\,A_{-},\,v_{f\,\bot} =\, 0 }\,.
\eeq
Taking into account the asymptotically leading contributions of $g^2$ order, that means the $M_{L}$ term presence in the expressions, see \cite{Kovner, Our2}, we obtain:
\beq\label{C3}
-2\,\imath\, K^{a\,b}_{x\,y\,1}\,=\,
\left[ G_0\,\frac{\delta^{2}\,M_{L}}{\delta A_{+\,x}^{a} \delta A_{-\,y}^{b}} -
G_0\,\frac{\delta\,M_{L}}{\delta A_{-\,y}^{b}}G_0\,\frac{\delta\,M_1}{\delta A_{+\,x}^{a}}-
G_0\,\frac{\delta\,M_{1}}{\delta A_{-\,y}^{b}}G_0\,\frac{\delta\,M_L}{\delta A_{+\,x}^{a}}
\right]_{A_{+},\,A_{-},\,v_{f\,\bot} =\, 0 }\,.
\eeq
For the first term we have:
\beq\label{C4}
-2\,\imath\, K^{a\,b}_{x\,y\,1,\,1}\,=\,G_0\,\frac{\delta^{2}\,M_{L}}{\delta A_{+\,x}^{a} \delta A_{-\,y}^{b}}\,=\,
G_{0\, + +}^{t z}\,\frac{g}{N}\,
\frac{\delta\,\Le U_{1}^{c d c}\Ra^{+}_{z t}}{\delta A_{+\,x}^{a}}\,
\frac{\delta\,  \D_{i}^{2} A_{-\,z}^{d}}{\delta A_{-\,y}^{b}}\,
\eeq
where the following identity was used:
\beq\label{C5}
\D_{i}\,\D_{-}\,\rho_{a}^{i}\,=\,-\,\frac{1}{N}\,\D_{i}^{2}\,A_{-}^{a}\,,
\eeq
see \eq{Ef5} and \cite{Our2}.
Using the following expressions
\beq\label{C6}
\frac{\delta\,\Le U_{1}^{c d c}\Ra^{+}_{z t}}{\delta A_{+\,x}^{a}}\,=\,g\,
\Le U_{2}^{c d c a_1}\Ra^{+ +}_{z t w}\,\frac{\delta\, v_{+\,w}^{a_1\,cl}}{\delta\,A_{+\,x}^{a}}\,
\eeq
and
\beq\label{C7}
\frac{\delta\, v_{+\,w}^{a_1\,cl}}{\delta\,A_{+\,x}^{a}}\,=\,\delta^{a\,a_1}\,\Le \delta^{2}_{x_{\bot}\,w_{\bot}} \delta_{x^{+}\,w^{+}}\Ra\,
\eeq
to requested accuracy, we obtain for \eq{C4}:
\beq\label{C8}
-2\,\imath\, K^{a\,b}_{x\,y\,1,\,1}\,=\,\frac{g^2}{N}\,G_{0\, + +}^{t z}\,\Le U_{2}^{c b c a_1}\Ra^{+ +}_{z t w}\,
\Le \delta^{a\,a_1}\,\delta^{2}_{x_{\bot}\,w_{\bot}} \delta_{x^{+}\,w^{+}}\,  \Ra\,
\Le \delta^{2}_{y_{\bot}\,z_{\bot}} \delta_{y^{-}\,z^{-}} \D_{i\,z}^{2}\,\Ra\,,
\eeq
where the NNLO term of the Lipatov's current series expansion reads as
\beqar\label{C9}
\Le\Le U_{2}^{c bc a}\Ra^{+ +}_{z t w}\,\Ra_{A_{+},\,A_{-} = \, 0}\,& = &\,\frac{1}{2}\,N^{2}\,\delta^{a b}\,\left[\,\Le\,
G_{z w}^{+\,0}\,G_{w t}^{+\,0}\,+\,G_{t w}^{+\,0}\,G_{w z}^{+\,0}\,\Ra\,+\,\right.\nonumber \\
&+&\,\left.
\,2\,\Le
G_{z t}^{+\,0}\,G_{t w}^{+\,0}\,+\,G_{z t}^{+\,0}\,G_{w z}^{+\,0}\,+\,G_{z w}^{+\,0}\,G_{t z}^{+\,0}\,+\,G_{t z}^{+\,0}\,G_{w t}^{+\,0}\,
\Ra\,\right]\,.
\eeqar
Therefore, writing explicitly all integrations in the expression, we obtain:
\beqar\label{C10}
&\,& -\,2\,\imath\, K^{a\,b}_{x\,y\,1,\,1}\,  = 
 \frac{1}{2}\, g^{2}\,N\,\delta^{a\,b}\,\int d^{4} z\,d^{4} t\, d^{4} w\,\Le\,\Le \D_{i\,z}^{2}\,G_{0\,+ +}^{t z}\Ra\, 
\Le \delta^{2}_{x_{\bot}\,w_{\bot}} \delta_{x^{+}\,w^{+}}\Ra\,\Le \delta^{2}_{y_{\bot}\,z_{\bot}} \delta_{y^{-}\,z^{-}}\Ra\,\right.\cdot\nonumber \\
&\cdot& \left. \left[\Le
G_{z w}^{+\,0}\,G_{w t}^{+\,0}+G_{t w}^{+\,0}\,G_{w z}^{+\,0}\Ra+
2\Le
G_{z t}^{+\,0}\,G_{t w}^{+\,0}+G_{z t}^{+\,0}\,G_{w z}^{+\,0}+G_{z w}^{+\,0}\,G_{t z}^{+\,0}+G_{t z}^{+\,0}\,G_{w t}^{+\,0}
\Ra\right] \Ra.
\eeqar

 Formally, there are three additional terms are present in \eq{C2}. The first one
\beq\label{C11}
-2\,\imath\, K^{a\,b}_{x\,y\,1,\,2}\,=\,-\,G_{0\, + +}\,\frac{\delta\,M_{L}}{\delta A_{-\,y}^{b}}\,\,G_{0\, + i}\,
\frac{\delta\,M_{1\,i -}}{\delta A_{+\,x}^{a}}\,,
\eeq
the second one
\beq\label{C12}
-2\,\imath\, K^{a\,b}_{x\,y\,1,\,3}\,=\,-\,G_{0\, i +}\,\frac{\delta\,M_{L}}{\delta A_{-\,y}^{b}}\,\,G_{0\, + +}\,
\frac{\delta\,M_{1\,- i}}{\delta A_{+\,x}^{a}}\,
\eeq
and the third one
\beq\label{C13}
-2\,\imath\, K^{a\,b}_{x\,y\,1,\,4}\,=\,-\,G_{0\, i +}\,\frac{\delta\,M_{L}}{\delta A_{-\,y}^{b}}\,G_{0\, + j}\,
\frac{\delta\,M_{1\,j i}}{\delta A_{+\,x}^{a}}\,.
\eeq
Nevertheless, only the third one contributes to the kernel in the limit of zero Reggeon fields, we have there: 
\beq\label{C14}
\frac{\delta\,M_{L}^{c d}}{\delta A_{-\,y}^{b}}\,=\,\frac{g}{N}\,
\Le U_{1}^{c b d}\Ra^{+}_{t w}\,\Le \delta^{2}_{y_{\bot}\,t_{\bot}} \delta_{y^{-}\,t^{-}} \D_{i\,t}^{2}\,\Ra\,
\eeq
where
\beq\label{C15}
\Le\Le U_{1}^{c b d}\Ra^{+}_{t w}\Ra_{A_{+},\,A_{-},\,v_{f\,\bot} =\, 0 }\,=\,\frac{1}{2}\,N\,f_{c d b}\,\Le G^{+\,0}_{t w}\,-\,G^{+\,0}_{w t} \Ra\,.
\eeq
Also we have:
\beq\label{C16}
\frac{\delta\,M_{j i}^{d c}}{\delta A_{+\,x}^{a}}\,=\,2\,g\,f_{d a c}\,\delta_{j\,i}\,\delta^{2}_{z_{\bot}\,x_{\bot}}\,\delta_{z^{+}\,x^{+}}\,\D_{-\,z}\,.
\eeq
The final expresion for this terms reads, therefore, as:
\beq\label{C17}
-2\imath\, K^{a\,b}_{x\,y\,1,\,4}=-\,g^2\,N\,\delta^{a b}\,\int\,d^4t\, d^4w\, d^4 z \,
\Le G^{+\,0}_{t w}-G^{+\,0}_{w t} \Ra\,\delta^{2}_{z_{\bot}\,x_{\bot}}\, \delta_{z^{+}\,x^{+}}\,\delta^{2}_{y_{\bot}\,t_{\bot}}\, \delta_{y^{-}\,t^{-}} \,
\Le\,G_{0\, + i}^{w z}\,\D_{-\,z}\,\D_{i\,t}^{2}\,G_{0\, i +}^{z t}\Ra\,.
\eeq
We notice that both \eq{C17} and \eq{C10} contributions are precisely the same as obtained in \cite{Our2} paper. Therefore, we immediately write the full contribution
from \cite{Our2} which is
\beq\label{C18}
K^{a\,b}_{x\,y\,1}\,= \,-\,\frac{\,g^{2}\,N}{8\,\pi}\,\D_{i\,x}^{2}\,\Le\,\int\,\frac{d p_{-}}{p_{-}}\,\int\,\frac{d^{2} p_{\bot}}{(2 \pi)^{2}}\,\int\,\frac{d^{2} k_{\bot}}{(2 \pi)^{2}}\,
\frac{ \,k_{\bot}^{2}}{p_{\bot}^{2}\,\Le\,p_{\bot}\,-\,k_{\bot}\,\Ra^{2}}\,e^{-\imath\,\,k_{i} \,\Le x^{i}\,-\,y^{i}\Ra}\,\Ra\,.
\eeq
We can rewrite this expression redefining the vertex as
\beq\label{C19}
K^{a\,b}_{x\,y\,1}\,\rightarrow\,K^{a\,b}_{x\,y\,1}\,\D_{i\,x}^{2}\,= \,\int\,\frac{d^{2} p}{(2 \pi)^2}\,\tilde{K}(p)\,e^{-\imath\,p_i\,\Le x^{i}\, - \,y^{i} \Ra}\,\D_{i\,x}^{2}\,
\eeq
with
\beq\label{C20}
\tilde{K}(p \,,\,\eta)\,=\,-\,\frac{\,N\,\pi\,g^{2}}{2}\,\delta(p_{\,+})\,\delta(p_{\,-})\,
\int_{0}^{\eta} d \eta^{'}\,\int \frac{d^2 k_{\bot}}{(2 \pi)^2}\,\frac{p_{\bot}^2}{k_{\bot}^{2}\,\Le p_{\bot} - k_{\bot} \Ra^2}\,,
\eeq
where the physical cut-off $\eta$ in rapidity space $y\,=\,\frac{1}{2} \ln(\Lambda\,k_-)$ is introduced.
Now, introducing the bare propagator of the Reggeon as
\beq\label{Pro61}
D_{+ -\, 0}^{a b}(x_{\bot},\,y_{\bot})\,=\,D_{0}^{a b}(x_{\bot},\,y_{\bot})\,=\,\delta^{a b}\,\int\,\frac{d^{2} p}{(2\pi)^2}\frac{e^{-\imath\,p_{i}\,(x^{i}\,-\,y^{i})}}{p_{\bot}^{2}}\,,
\eeq
the full propagator to the leading order precision can be written as the following equation
\beq\label{Pro62}
 D_{x y}^{a c}\,=\, D_{x y\,0}^{a c}\,-\,\int\,d^{4} z\,\int\,d^{4} w\,\Le \D_{i\,z}^{2} D_{x z\,0}^{a b}\,\Ra\,
K^{b d}_{z w\,1}\, D_{w y}^{d c}\,.
\eeq
Introducing
\beq\label{Pro7}
 D_{x y}^{a c}\,=\,\delta^{a c}\delta(y^{-}\,-\,x^{-})\,\delta(x^{+}\,-\,y^{+})\,\int\,\frac{d^2 p}{(2 \pi)^2}\,\tilde{D}(p_{\bot},\,\eta)\,
e^{-\imath\,p_{i}\,\Le x^{i} - y^{i}\Ra}\,,
\eeq
we obtain finally:
\beq\label{Pro8}
\tilde{D}^{a b}(p_{\bot},\eta)\,=\,\frac{\delta^{a b}}{p_{\bot}^{2}}\,+\,
\epsilon(p_{\bot}^{2})\,\int^{\eta}_{0}\,d\eta^{'}\,\tilde{D}^{a b}(p_{\bot},\eta^{'})\,
\eeq
with
\beq\label{Pro9}
\epsilon(p_{\bot}^{2})\,=\,-\,\frac{\alpha_{s}\,N}{4\,\pi^{2}}\,\int\,d^{2} k_{\bot}\,
\frac{ \,p_{\bot}^{2}}{k_{\bot}^{2}\,\Le\,p_{\bot}\,-\,k_{\bot}\,\Ra^{2}}\,
\eeq
as trajectory of the propagator of reggeized gluons. Rewriting this equation as the differential one:
\beq\label{Pro10}
\frac{\D\,\tilde{D}^{a b}(p_{\bot},\eta)}{\D\,\eta}\,=\,\tilde{D}^{a b}(p_{\bot},\eta)\,\epsilon(p_{\bot}^{2})\,
\eeq
we obtain the finally the propagator:
\beq\label{Pro11}
\tilde{D}^{a b}(p_{\bot},\eta)\,=\,\frac{\delta^{a b}}{p_{\bot}^{2}}\,e^{\,\eta\,\epsilon(p_{\bot}^{2})}\,,
\eeq
with $\eta$ defined in some rapidity interval $0\,<\,\eta\,<\,Y\,=\,\ln(s/s_{0})$ of interest; it is the BFKL propagator for reggeized gluons, see \cite{BFKL}.

\newpage
\section*{Appendix D: NLO vertex of interactions of $A_{+}\,A_{+}\,A_{-}$ Reggeon fields}
\renewcommand{\theequation}{D.\arabic{equation}}
\setcounter{equation}{0}

  In the following calculations we omit the production fields in the expressions taking them equal to zero from the beginning. We also do not denote but mean 
zero limit of the all  Reggeon fields in the end of the functional derivatives calculations, also the trace in the expressions is assumed. 
Therefore, for the NLO (one loop or $g^3$ order) vertex of interaction of 
$A_{+}\,A_{+}\,A_{-}$ Reggeon fields we have:
\beqar\label{D1}
&-&2\,\imath\, K^{a\,b\,c}_{x\,y\,z\,1}\,=\,\frac{\delta^{3}\,\ln\Le 1 + G_0\,M\,\Ra}{\delta A_{+\,x}^{a}\,\delta A_{-\,y}^{b}\,A_{+\,z}^{c}}\,=\,
G_0\,\frac{\delta^{3}\,M}{\delta A_{+\,x}^{a} \delta A_{-\,y}^{b} \delta A_{+\,z}^{c}} \,-\,
G_0\,\frac{\delta^{2}\,M}{\delta A_{+\,x}^{a} \delta A_{-\,y}^{b}}\, G_{0}\, \frac{\delta\,M}{\delta A_{+\,z}^{c}} \,-\,\nonumber \\
&-& \,G_0\,\frac{\delta^{2}\,M}{\delta A_{+\,z}^{c} \delta A_{-\,y}^{b}}\, G_{0}\, \frac{\delta\,M}{\delta A_{+\,x}^{a}} \,-\,
G_0\,\frac{\delta^{2}\,M}{\delta A_{+\,x}^{a} \delta A_{+\,z}^{c}}\, G_{0}\, \frac{\delta\,M}{\delta A_{-\,y}^{b}}\, +\,
G_0\,\frac{\delta\,M}{\delta A_{-\,y}^{b}}\, G_0\,\frac{\delta\,M}{\delta A_{+\,z}^{c}}\,G_{0}\, \frac{\delta\,M}{\delta A_{+\,x}^{a}}\,+
\nonumber \\
&+&
G_0\,\frac{\delta\,M}{\delta A_{-\,y}^{b}}\, G_0\,\frac{\delta\,M}{\delta A_{+\,x}^{a}}\,G_{0}\, \frac{\delta\,M}{\delta A_{+\,z}^{c}}\,.
\eeqar
Similarly to the done in the previous Appendix, we keep in the  \eq{D1} expression only terms which are arising from the $M_{L}$ term in \eq{Ef2} and which provide leading asymptotic 
contributions: 
\beqar\label{D2}
&-&2\,\imath\, K^{a\,b\,c}_{x\,y\,z\,1}\,=\,\frac{\delta^{3}\,\ln\Le 1 + G_0\,M\,\Ra}{\delta A_{+\,x}^{a}\,\delta A_{-\,y}^{b}\,A_{+\,z}^{c}}\,=\,
G_0\,\frac{\delta^{3}\,M_{L}}{\delta A_{+\,x}^{a} \delta A_{-\,y}^{b} \delta A_{+\,z}^{c}} \,-\,
G_0\,\frac{\delta^{2}\,M_{L}}{\delta A_{+\,x}^{a} \delta A_{-\,y}^{b}}\, G_{0}\, \frac{\delta\,M_{1}}{\delta A_{+\,z}^{c}} \,-\,\nonumber \\
&-& \,G_0\,\frac{\delta^{2}\,M_{L}}{\delta A_{+\,z}^{c} \delta A_{-\,y}^{b}}\, G_{0}\, \frac{\delta\,M_{1}}{\delta A_{+\,x}^{a}} \,+\,
G_0\,\frac{\delta\,M_{L}}{\delta A_{-\,y}^{b}}\, G_0\,\frac{\delta\,M_{1}}{\delta A_{+\,z}^{c}}\,G_{0}\, \frac{\delta\,M_{1}}{\delta A_{+\,x}^{a}}\,+
\nonumber \\
&+&
G_0\,\frac{\delta\,M_{L}}{\delta A_{-\,y}^{b}}\, G_0\,\frac{\delta\,M_{1}}{\delta A_{+\,x}^{a}}\,G_{0}\, \frac{\delta\,M_{1}}{\delta A_{+\,z}^{c}}\,.
\eeqar
Further we consider all relavant terms one by one.

\subsection*{First contribution}

For the first term in the r.h.s. of \eq{D2} we obtain:
\beq\label{D3}
-2\,\imath\, K^{a\,b\,c}_{x\,y\,z\,1,\,1}\,=\,G_0\,\frac{\delta^{3}\,M_{L}}{\delta A_{+\,x}^{a} \delta A_{-\,y}^{b}  A_{+\,z}^{c} }\,=\,
G_{0\, + +}^{t p}\,\frac{g}{N}\,
\frac{\delta^{2}\,\Le U_{1}^{k d k}\Ra^{+}_{p t}}{\delta A_{+\,x}^{a} \delta A_{+\,z}^{c}}\,
\frac{\delta\,  \D_{i}^{2} A_{-\,p}^{d}}{\delta A_{-\,y}^{b}}\,,
\eeq
see \eq{C4} in the previous Appendix. 
Now, using \eq{C6} expression,  we have:
\beq\label{D4}
\frac{\delta^2\,\Le U_{1}^{b d b}\Ra^{+}_{p t}}{\delta A_{+\,x}^{a}\,\delta A_{+\,z}^{c}}\,=\,g\,
\frac{\delta \Le U_{2}^{b d b a_1}\Ra^{+ +}_{p t w}}{\delta  A_{+\,z}^{c}}\,\frac{\delta\, v_{+\,w}^{a_1\,cl}}{\delta\,A_{+\,x}^{a}}\,+\,g\,
\Le U_{2}^{b d b a_1}\Ra^{+ +}_{p t w}\,\frac{\delta^2\, v_{+\,w}^{a_1\,cl}}{\delta\,A_{+\,x}^{a} \delta A_{+\,z}^{c}}\,.
\eeq
The only first term of \eq{D4} will remain in the limit of the zero Reggeon fields ,
see the NLO value of $v_{+}^{cl}$ in \cite{Our2,Our41},
therefore 
\beq\label{D41}
\frac{\delta^2\,\Le U_{1}^{b d b}\Ra^{+}_{p t}}{\delta A_{+\,x}^{a}\,\delta A_{+\,z}^{c}}\,=\,g^2\,\Le U_{3}^{b d b a_{1} a_{2}}\Ra^{+ + +}_{p t w w_1}\,
\frac{\delta\, v_{+\,w}^{a_1\,cl}}{\delta\,A_{+\,x}^{a}}\,\frac{\delta\, v_{+\,w_1}^{a_2\,cl}}{\delta\,A_{+\,z}^{c}}\,,
\eeq
where in the expression
\beq\label{D5}
\Le U_{3}^{a b c d e}\Ra^{+ + +}\,=\,\sum\,tr\,[f_{\{a\,}\,G^{+} \,f_{e}\,G^{+}\,f_{c}\,G^{+}\,f_{d\}}\,O\,f_{b}\,O^{T}]
\eeq
the sum is performed on the permutations of the $a\, c\, d\, e\,$ indexes, see \eq{Ef4} and Appendix A for the derivation of the expression.
There are the following traces of the color matrices in the adjoint representation we need to know. The first one is the following one:
\beqar\label{D6}
C^{5}_{c d c a b}&=&tr[f_c f_d f_c f_a f_b]\,=\,tr[f_d f_c f_c f_a f_b]\,-\,f_{c d c_1}\,tr[f_{c_1} f_c f_a f_b]\,=\,-N\,
tr[f_d f_a f_b]\,-\,f_{c d c_1}\,tr[f_{c_1} f_c f_a f_b]\,=\,\nonumber\\
&=&\,-\,\frac{N^2}{2}\,f_{d a b}\,-\,f_{c d c_1}\,\Le \delta_{c_1 b}\delta_{a c} \,+\,
\frac{1}{2}\Le \delta_{c_1 c}\delta_{a b} + \delta_{c_1 a}\delta_{c b}\Ra \,+\,
\frac{N}{4}\Le f_{c_1 b e} f_{e c a} + d_{c_1 b e} d_{e c a} \Ra\,\Ra\,=\,\nonumber \\
&=&\,\frac{N^2}{2}\,f_{a d b}\,-\,f_{a d b}\,+\,\frac{1}{2}\,f_{a d b}\,-\,\frac{N}{4}\,f_{c d c_1}\,f_{c_1 b e} f_{e c a}\,-\,
\frac{N}{4}\,f_{c d c_1}\,d_{c_1 b e} d_{e c a}\,,
\eeqar
here $d$ is fully symmetric tensor:
\beq\label{D7}
d_{a b c}\,=\,2\,tr[\{T^a,\, T^{b}\}\,T^{c}]
\eeq
with $T^{a}$ as a color matrix in a fundamental representation. 
Now, using the following identities:  
\beq\label{D8}
f_{d c_1 c}\, f_{c_1 b e}\,f_{c  e a}\,=\,-\,\frac{1}{2}\,N\,f_{d b a}
\eeq
and
\beq\label{D9}
f_{d c_1 c}\,d_{c_1 b e}\,d_{c e a }\,=\,\Le \frac{N^2\,-\,4}{2 N} \Ra\,f_{d b a}\,,
\eeq
see \cite{Fadin}, we obtain finally for the factor:
\beq\label{D10}
C^{5}_{c d c a b}\,=\,tr[f_c f_d f_c f_a f_b]\,=\, \frac{N^2}{4}\,f_{a d b}\,.
\eeq
An another color factor we need is the following one:
\beq\label{D11}
C^{5}_{c c d a b}\,=\,tr[f_c f_c f_d f_a f_b]\,=\,\frac{N^2}{2}\,f_{a d b}\,. 
\eeq
Therefore, we obtain for the \eq{D5} expression:
\beqar\label{D12}
&\,&\Le U_{3}^{b d b a_1 a_2}\Ra^{+ + +}_{p t w w_1}\,=\,C^{5}_{b a_2 b a_1 d}\,G^{+\,0}_{p w_1}\,G^{+\,0}_{w_1 t}\,G^{+\,0}_{t w}\,+\,
C^{5}_{b b a_2  a_1 d}\,\Le\,G^{+\,0}_{p t}\,G^{+\,0}_{t w_1}\,G^{+\,0}_{w_1 w}\,-\,
G^{+\,0}_{p t}\,G^{+\,0}_{t w}\,G^{+\,0}_{w w_1}\,+\,
\right.\nonumber \\ 
&+&\,\left.\,G^{+\,0}_{w_1 p}\,G^{+\,0}_{p t}\,G^{+\,0}_{t w}\,\Ra\,+\,C^{5}_{b a_2 b a_1 d}\,\Le
\,G^{+\,0}_{p w_1}\,G^{+\,0}_{w_1 w}\,G^{+\,0}_{w t}\,-\,G^{+\,0}_{p w}\,G^{+\,0}_{w w_1}\,G^{+\,0}_{w_1 t}\,-\,
\,G^{+\,0}_{p w}\,G^{+\,0}_{w t}\,G^{+\,0}_{t w_1}\,+\,
\right.\nonumber \\
&+&\,\left. G^{+\,0}_{w_1 p}\,G^{+\,0}_{p w}\,G^{+\,0}_{w t}\,\Ra\,-\,C^{5}_{b a_2 b a_1 d}\,G^{+\,0}_{w p}\,G^{+\,0}_{p w_1}\,G^{+\,0}_{w_1 t}\,+\,
C^{5}_{b b a_2  a_1 d}\,\Le\,G^{+\,0}_{w_1 w}\,G^{+\,0}_{w p}\,G^{+\,0}_{p t}\,-\,\,G^{+\,0}_{w w_1}\,G^{+\,0}_{w_1 p}\,G^{+\,0}_{p t}\,-\,
\right. \nonumber\\
&-&\,\left. \,G^{+\,0}_{w p}\,G^{+\,0}_{p t}\,G^{+\,0}_{t w_1}\, \Ra\,+\,
C^{5}_{b a_2 b a_1 d}\,G^{+\,0}_{t w_1}\,G^{+\,0}_{w_1 p}\,G^{+\,0}_{p w}\,+\,
C^{5}_{b b a_2  a_1 d}\,\Le\,G^{+\,0}_{t p}\,G^{+\,0}_{p w_1}\,G^{+\,0}_{w_1 w}\,-\,
G^{+\,0}_{t p}\,G^{+\,0}_{p w}\,G^{+\,0}_{w w_1}\,+\,
\right.\nonumber \\ 
&+&\,\left.\,G^{+\,0}_{w_1 t}\,G^{+\,0}_{t p}\,G^{+\,0}_{p w}\,\Ra\,-\,C^{5}_{b a_2 b a_1 d}\,G^{+\,0}_{w t}\,G^{+\,0}_{t w_1}\,G^{+\,0}_{w_1 p}\,+\,
C^{5}_{b b a_2  a_1 d}\,\Le\,G^{+\,0}_{w_1 w}\,G^{+\,0}_{w t}\,G^{+\,0}_{t p}\,-\,\,G^{+\,0}_{w w_1}\,G^{+\,0}_{w_1 t}\,G^{+\,0}_{t p}\,-\,
\right. \nonumber\\
&-&\,\left. \,G^{+\,0}_{w t}\,G^{+\,0}_{t p}\,G^{+\,0}_{p w_1}\, \Ra\,+\,
\,C^{5}_{b a_2 b a_1 d}\,\Le
\,G^{+\,0}_{t w_1}\,G^{+\,0}_{w_1 w}\,G^{+\,0}_{w p}\,-\,G^{+\,0}_{t w}\,G^{+\,0}_{w w_1}\,G^{+\,0}_{w_1 p}\,-\,
\,G^{+\,0}_{t w}\,G^{+\,0}_{w p}\,G^{+\,0}_{p w_1}\,+\,
\right.\nonumber \\
&+&\,\left. G^{+\,0}_{w_1 t}\,G^{+\,0}_{t w}\,G^{+\,0}_{w p}\,\Ra\,,
\eeqar
or
\beqar\label{D13}
&\,&\Le U_{3}^{b d b a_1 a_2}\Ra^{+ + +}_{p t w w_1}\,=\,
C^{5}_{b b a_2  a_1 d}\,\Le\,
G^{+\,0}_{p t}\,G^{+\,0}_{t w_1}\,G^{+\,0}_{w_1 w}\,-\,G^{+\,0}_{p t}\,G^{+\,0}_{t w}\,G^{+\,0}_{w w_1}\,+\,G^{+\,0}_{w_1 p}\,G^{+\,0}_{p t}\,G^{+\,0}_{t w}\,+\,
\right.\nonumber \\ 
&+&\,\left.
\,G^{+\,0}_{w_1 w}\,G^{+\,0}_{w p}\,G^{+\,0}_{p t}\,-\,\,G^{+\,0}_{w w_1}\,G^{+\,0}_{w_1 p}\,G^{+\,0}_{p t}\,-\,G^{+\,0}_{w p}\,G^{+\,0}_{p t}\,G^{+\,0}_{t w_1}\,+\,
\right.\nonumber \\ 
&+&\,\left.
\,G^{+\,0}_{t p}\,G^{+\,0}_{p w_1}\,G^{+\,0}_{w_1 w}\,-\,G^{+\,0}_{t p}\,G^{+\,0}_{p w}\,G^{+\,0}_{w w_1}\,+\,G^{+\,0}_{w_1 t}\,G^{+\,0}_{t p}\,G^{+\,0}_{p w}\,+\,
\right.\nonumber \\ 
&+&\,\left.
\,G^{+\,0}_{w_1 w}\,G^{+\,0}_{w t}\,G^{+\,0}_{t p}\,-\,\,G^{+\,0}_{w w_1}\,G^{+\,0}_{w_1 t}\,G^{+\,0}_{t p}\,-\,G^{+\,0}_{w t}\,G^{+\,0}_{t p}\,G^{+\,0}_{p w_1}\,\Ra\,+\,
\nonumber \\ 
&+&\,
C^{5}_{b a_2 b a_1 d}\,\Le \, G^{+\,0}_{p w_1}\,G^{+\,0}_{w_1 t}\,G^{+\,0}_{t w}\,+\,
\,G^{+\,0}_{p w_1}\,G^{+\,0}_{w_1 w}\,G^{+\,0}_{w t}\,-\,G^{+\,0}_{p w}\,G^{+\,0}_{w w_1}\,G^{+\,0}_{w_1 t}\,-\,
\,G^{+\,0}_{p w}\,G^{+\,0}_{w t}\,G^{+\,0}_{t w_1}\,+\,
\right.\nonumber \\
&+&\,\left.
G^{+\,0}_{w_1 p}\,G^{+\,0}_{p w}\,G^{+\,0}_{w t}\,-\,G^{+\,0}_{w p}\,G^{+\,0}_{p w_1}\,G^{+\,0}_{w_1 t}\,+\,G^{+\,0}_{t w_1}\,G^{+\,0}_{w_1 p}\,G^{+\,0}_{p w}\,-\,
G^{+\,0}_{w t}\,G^{+\,0}_{t w_1}\,G^{+\,0}_{w_1 p}\,+\,
\right.\nonumber \\
&+&\,\left.
\,G^{+\,0}_{t w_1}\,G^{+\,0}_{w_1 w}\,G^{+\,0}_{w p}\,-\,G^{+\,0}_{t w}\,G^{+\,0}_{w w_1}\,G^{+\,0}_{w_1 p}\,-\,
\,G^{+\,0}_{t w}\,G^{+\,0}_{w p}\,G^{+\,0}_{p w_1}\,+\,G^{+\,0}_{w_1 t}\,G^{+\,0}_{t w}\,G^{+\,0}_{w p}\,\Ra\,.
\eeqar
In order to reduce the complexity of the calculations we note the following. For the symmetrical with respect to $k_-$ (rapidity) contributions we can restrict the integrals only by positive values of $k_-$.
Namely, regularizing the integral over the $k_-$ we in general obtain the following type of expressions
\beq\label{D1311}
\int \frac{d k_-}{k_-}\,I\rightarrow\, \Le \int_{\epsilon}^{1/\epsilon} \frac{d k_-}{k_-}\,I_1\,+\,\int_{- 1/\epsilon}^{-\epsilon} \frac{d k_-}{k_-}\,I_2\Ra\,,
\eeq
where $I_i$ is a corresponding expressions obtained by integration in respect to other coordinates. Now, taking into account that for the large part of the contributions the following condition holds
\beq\label{D1312} 
I_1 = -I_2\,,
\eeq
which is related to the different directions of the integration contours in the complex plane of $k_{+}$ variable, 
we use the following regularization of $k_-$ integrals:
\beq\label{D1313} 
\int \frac{d k_-}{k_-}\,I\rightarrow\,\frac{1}{2} \Le \int_{\epsilon}^{1/\epsilon} \frac{d k_-}{k_-}\,-\,\int_{- 1/\epsilon}^{-\epsilon} \frac{d k_-}{k_-}\,\Ra\,I_1 =\,\int_{\Lambda}^{k_{-}^{ max}} \frac{d k_-}{k_-}\,I_1\,.
\eeq
Therefore, in general, for the contributions which are symmetrical with respect to the $k_{-}$ momenta in the sense of \eq{D1312}, we can restrict the integrals over $k_-$ only by positive values introducing the 
rapidity variable as $y\,=\,\frac{1}{2}\,\ln\Le \frac{k_-}{\Lambda}\Ra$ and obtaining
\beq\label{D1314}
\int \frac{d k_-}{k_-}\,I\,=\,\int_{\Lambda}^{k_{-}^{ max}} \frac{d k_-}{k_-}\,I_{1}\,=\,2\,\int_{0}^{\eta} dy \,I_1
\eeq
with $\eta$ as ultraviolet cut-off related to the value of the particle's cluster in the effective action approach.
Consequently, we notice that the
integrals which consist of $\theta^{\,+}_{p t}$ function and have no any singularities in integration with respect to the $+$
components of the coordinates are zero due the fact that
\beq\label{D14}
G_{0\,++}\,\propto\,\int\,d k_{+}\,\frac{e^{-\,\imath\,k_{+}\,\Le\, t^{+}\,-\,p^{+}\,\Ra}}{k_{+}\,-\,k_{\bot}^{2}/2\,k_{-}\,+\,\imath\,\varepsilon}\,\propto\,\theta(t^{+}\,-\,p^{+})\,.
\eeq
There are the following terms which are zero because of that reason: $G^{+\,0}_{p w_1}\,G^{+\,0}_{w_1 t}\,G^{+\,0}_{t w}\,$, 
$G^{+\,0}_{p w_1}\,G^{+\,0}_{w_1 w}\,G^{+\,0}_{w t}$ and $G^{+\,0}_{w_1 p}\,G^{+\,0}_{p w}\,G^{+\,0}_{w t}$ and corresponding terms obtained after the 
$ w\,\rightleftarrows\,w_1 $ substitution. We also note, that after the integration with respect to all delta functions in \eq{D3} and $G^{+\,0}$ functions,
the remaining answer will depend only on transverse $\delta^{2}_{y_{\bot}\,x_{\bot}}$ and $\delta^{2}_{y_{\bot}\,z_{\bot}}\,$ functions. Therefore, further, for the
shortening of notations, we will use $G^{+\,0}$ functions as if it equivalent to the theta functions, remembering that all delta functions are remain in the final expressions with any number of 
theta functions  ($G^{+\,0}$ functions)  in them.
There are the following remaining terms we have to account:
\beqar\label{D142}
&\,&\Le U_{3}^{b d b a_1 a_2}\Ra^{+ + +}_{p t w w_1}\,=\,-\,f_{a_1 d a_2}\,
\frac{N^2}{4}\,\Le\,
2\,\Le G^{+\,0}_{p t}\,G^{+\,0}_{t w_1}+\,\,G^{+\,0}_{t p}\,G^{+\,0}_{p w_1}\Ra\,G^{+\,0}_{w_1 w}\,+\,
2\,G^{+\,0}_{w_1 w}\,\Le G^{+\,0}_{w p}\,G^{+\,0}_{p t}\,+\,
\right.\right. \nonumber \\
&+&\,\left. \left.
G^{+\,0}_{w t}\,G^{+\,0}_{t p}\Ra\,+
2\,\Le G^{+\,0}_{w_1 t}\,-\,G^{+\,0}_{w_1 t}\,G^{+\,0}_{t p}\,-\,G^{+\,0}_{p w_1}\,G^{+\,0}_{w_1 t} \Ra\,G^{+\,0}_{t w}\,
+\,\right. \nonumber \\
&+&\,\left.
2\,\Le G^{+\,0}_{w_1 p}\,-\,G^{+\,0}_{w_1 p}\,G^{+\,0}_{p t}\,-\,G^{+\,0}_{t w_1}\,G^{+\,0}_{w_1 p} \Ra\,G^{+\,0}_{p w}\,+\,\right. \nonumber \\
&+&\left.\,G^{+\,0}_{t w_1}\,G^{+\,0}_{w_1 p}\,G^{+\,0}_{p w}\,+\,G^{+\,0}_{t w_1}\,G^{+\,0}_{w_1 w}\,G^{+\,0}_{w p}\,+\,
G^{+\,0}_{w_1 t}\,G^{+\,0}_{t w}\,G^{+\,0}_{w p}\,-\,\Le w\,\rightleftarrows\,w_1 \Ra\,\Ra\,.
\eeqar
We used and will use here the following identity for theta functions:
\beq\label{D143} 
\theta^{+}_{x y}\,\theta^{+}_{y z}\,=\, \theta^{+}_{x z}\,-\,\theta^{+}_{x z}\,\theta^{+}_{z y}\, -\,\theta^{+}_{y x}\,\theta^{+}_{x z}\,
\eeq
and correspondingly we obtain:
\beqar\label{D1421}
&\,&\Le U_{3}^{b d b a_1 a_2}\Ra^{+ + +}_{p t w w_1}\,=\,-\,f_{a_1 d a_2}\,
\frac{N^2}{4}\,\Le\,
2\,G^{+\,0}_{p w_1}\,G^{+\,0}_{w_1 w}\,-\,2\,\,G^{+\,0}_{p w_1}\,G^{+\,0}_{w_1 t}\,G^{+\,0}_{w_1 w}\,+\,
2\,G^{+\,0}_{w_1 w}\,G^{+\,0}_{w t}\,-\,
\right. \nonumber \\
&-&\,\left. 
2\,G^{+\,0}_{w_1 w}\,G^{+\,0}_{p w}\,G^{+\,0}_{w t}\,+\,
2\,\Le G^{+\,0}_{w_1 t}\,-\,G^{+\,0}_{w_1 t}\,G^{+\,0}_{t p}\,-\,G^{+\,0}_{p w_1}\,G^{+\,0}_{w_1 t} \Ra\,G^{+\,0}_{t w}\,+\,\right. \nonumber \\
&+&\,\left.
2\,\Le G^{+\,0}_{w_1 p}\,-\,G^{+\,0}_{w_1 p}\,G^{+\,0}_{p t}\,-\,G^{+\,0}_{t w_1}\,G^{+\,0}_{w_1 p} \Ra\,G^{+\,0}_{p w}\,+\,\right. \nonumber \\
&+&\left.\,G^{+\,0}_{t w_1}\,G^{+\,0}_{w_1 p}\,G^{+\,0}_{p w}\,+\,G^{+\,0}_{t w_1}\,G^{+\,0}_{w_1 w}\,G^{+\,0}_{w p}\,+\,
G^{+\,0}_{w_1 t}\,G^{+\,0}_{t w}\,G^{+\,0}_{w p}\,-\,\Le w\,\rightleftarrows\,w_1 \Ra\,\Ra\,.
\eeqar
Again using \eq{D143} identity and preserving only non-zero terms  we write \eq{D1421} in the following form:
\beqar\label{D145}
\Le U_{3}^{b d b a_1 a_2}\Ra^{+ + +}_{p t w w_1}\,& = &\,-\,f_{a_1 d a_2}\,
\frac{N^2}{4}\, 
\Le\, 2\,G^{+\,0}_{p w_1}\,G^{+\,0}_{w_1 w}\, +\,
2\,G^{+\,0}_{w_1 w}\,G^{+\,0}_{w t}\,+\,
2\,G^{+\,0}_{w_1 t}\,G^{+\,0}_{t w}\,+\,
2\,G^{+\,0}_{w_1 p}\,G^{+\,0}_{p w}\,-\,
\right.\nonumber \\
&-&\,\left.
2\,G^{+\,0}_{w_1 t}\,G^{+\,0}_{t p}\,G^{+\,0}_{t w}\,-\,
2\,G^{+\,0}_{t w_1}\,G^{+\,0}_{w_1 p}\,G^{+\,0}_{p w}\,+\,
G^{+\,0}_{w_1 w}\,G^{+\,0}_{w p}\,+\,
G^{+\,0}_{t w_1}\,G^{+\,0}_{w_1 w}\,-\,
\right. \nonumber \\
&-&\,\left.\,
G^{+\,0}_{t w_1}\,G^{+\,0}_{w_1 w}\,G^{+\,0}_{w p}\,-\,
G^{+\,0}_{w_1 w}\,G^{+\,0}_{w t}\,G^{+\,0}_{w p}\,-\,
G^{+\,0}_{t w_1}\,G^{+\,0}_{p w_1}\,G^{+\,0}_{w_1 w}-
\Le w\,\rightleftarrows\,w_1 \Ra\,\Ra\,.
\eeqar
The contributions of all integrals in \eq{D145} expression are proportional to the same tadpole integral and
we consider contributions of different terms in \eq{D145} one by one calculating the overall coefficient of the expression.
\begin{enumerate}
\item
The first four terms in \eq{D145} provide together zero contribution to \eq{D3}:
\beq\label{D146}
I_{1-4}\,=\,-\,16\,\pi\,\imath\,k_{-}\,\theta (w_1 - w)\,\int\,\frac{ d^{2} k_{\bot}}{k_{\bot}^{2}}\,,
\eeq
with only two first terms contributing to the answer. 
\item
The next term, $\,G^{+\,0}_{w_1 t}\,G^{+\,0}_{t p}\,G^{+\,0}_{t w}\,$ we rewrite as following:
\beq\label{D148}
G^{+\,0}_{w_1 t}\,G^{+\,0}_{t p}\,G^{+\,0}_{t w}\,=\,G^{+\,0}_{w_1 t}\,G^{+\,0}_{t w}\,-G^{+\,0}_{w_1 t}\,G^{+\,0}_{p t}\,G^{+\,0}_{t w}\,,
\eeq
the integration on $k_{+}$ variable gives zero for these contributions as well.
\item
The next term we consider is the $\,-\,2\,G^{+\,0}_{t w_1}\,G^{+\,0}_{w_1 p}\,G^{+\,0}_{p w}\,$ expression. We have after the integration on $+$
components of coordinates:
\beq\label{D18}
2\,\int\,d k_{+}\,\frac{1\,-\,e^{-\,\imath\,k_{+}\,\Le\, w^{+}_{1}\,-\,w^{+}\,\Ra}}{\Le\,k_{+}\,-\,\imath\,\varepsilon\,\Ra\,\Le\,
k_{+}\,-\,k_{\bot}^{2}/2\,k_{-}\,+\,\imath\,\varepsilon\Ra}\,=\,-\,4 \pi \imath\,\frac{2 k_-}{k_{\bot}^{2}}\,\Le
1\,-\,e^{-\,\imath\,\frac{k_{\bot}^{2}}{2\,k_{-}}\,(w_{1}^{+}\,-\,w^{+})} \Ra\,\theta(w_{1}^{+}\,-\,w^{+}).
\eeq
that, following to 't Hooft-Veltman conjecture, 
see \cite{Leibb}, gives zero final final contribution after an integration on transverse momenta.
\item
The terms $G^{+\,0}_{w_1 w}\,G^{+\,0}_{w p}$ and $
G^{+\,0}_{t w_1}\,G^{+\,0}_{w_1 w}$ after the integration on $k_{+}$ are equal to 
\beq\label{D181}
8\,\pi\,\imath\,k_{-}\,\theta (w_1 - w)\,\int\,\frac{ d^{2} k_{\bot}}{k_{\bot}^{2}}\,.
\eeq
\item
The $\,-\,G^{+\,0}_{t w_1}\,G^{+\,0}_{w_1 w}\,G^{+\,0}_{w p}\,$ expression in \eq{D145} is proportional to the following integral:
\beq\label{D188}
\int\,d k_{+}\,\frac{\,e^{-\,\imath\,k_{+}\,\Le\, w^{+}_{1}\,-\,w^{+}\,\Ra}}{\Le\,k_{+}\,-\,\imath\,\varepsilon\,\Ra\,\Le\,
k_{+}\,-\,k_{\bot}^{2}/2\,k_{-}\,+\,\imath\,\varepsilon\Ra}\,=\,
-\,2 \pi \imath\,\frac{2 k_-}{k_{\bot}^{2}}\,\theta (w_1 - w)\,e^{-\,\imath\,\frac{k_{\bot}^{2}}{2\,k_{-}}\,(w_{1}^{+}\,-\,w^{+})}\,,
\eeq
Expanding the exponential and keeping only non-zero in the sense of 't Hooft-Veltman conjecture terms, we obtain for this contribution:
\beq\label{D20}
\int\,d^{2} k_{\bot}\,\int\,d k_{+}\,\frac{e^{-\,\imath\,k_{+}\,\Le\, w^{+}_{1}\,-\,w^{+}\,\Ra}}{\Le\,k_{+}\,-\,\imath\,\varepsilon\,\Ra\,\Le\,
k_{+}\,-\,k_{\bot}^{2}/2\,k_{-}\,+\,\imath\,\varepsilon\Ra}\,=\,-\,4\,\pi\,\imath\,k_{-}\,\theta (w_1 - w)\,\int\,\frac{ d^{2} k_{\bot}}{k_{\bot}^{2}}\,.
\eeq
\item
The $G^{+\,0}_{w_1 w}\,G^{+\,0}_{w t}\,G^{+\,0}_{w p}$ and $G^{+\,0}_{t w_1}\,G^{+\,0}_{p w_1}\,G^{+\,0}_{w_1 w}$ terms, in turn, provides the following integral
\beq\label{D18111}
-2\,\int\,d^{2} k_{\bot}\,\int\,d k_{+}\,\frac{\theta (w_1 - w)\,k_{+}}{\Le\,k_{+}\,+\,\imath\,\varepsilon\,\Ra\,\Le\,k_{+}\,-\,\imath\,\varepsilon\,\Ra\,\Le\,
k_{+}\,-\,k_{\bot}^{2}/2\,k_{-}\,+\,\imath\,\varepsilon\Ra}\,=\,4\,\pi\,\imath\,k_{-}\,\theta (w_1 - w)\,\int\,\frac{ d^{2} k_{\bot}}{k_{\bot}^{2}}\,.
\eeq
\end{enumerate}

Finally, performing all requested integration, we obtain for the \eq{D3} contribution:
\beq\label{D21}
-2\imath\, K^{a\,b\,c}_{x\,y\,z\,1,\,1}=\,-\,\imath\,\frac{g^3\,N}{8\,\pi^3}\,f_{a b c}\,
\Le \theta (z^{+}-x^{+})\, - \,\theta (x^{+}-z^{+})\Ra\,\D_{\bot\,y}^{2}\Le
\delta^{2}_{y_{\bot}\,x_{\bot}}\,\delta^{2}_{y_{\bot}\,z_{\bot}}\,
\int\,\frac{d k_-}{k_-}\,\int\,\frac{d^{2} k_{\bot}}{k_{\bot}^{2}}\Ra\,.
\eeq

\subsection*{Second contribution}

 Now, consider the second term (the third one can be obtain from this term by $a,x\,\rightleftarrows\,c,z$ substitution) in \eq{D2}.
\beqar\label{D22}
-2\,\imath\, K^{a\,b\,c}_{x\,y\,z\,1,\,2}\,& = &\,-\,
G_{0\,+ +}\,\frac{\delta^{2}\,M_{L}^{+ +}}{\delta A_{+\,x}^{a} \delta A_{-\,y}^{b}}\, G_{0\,+ i}\, \frac{\delta\,M_{1}^{i +}}{\delta A_{+\,z}^{c}} \,
-\,G_{0\,i +}\,\frac{\delta^{2}\,M_{L}^{+ +}}{\delta A_{+\,x}^{a} \delta A_{-\,y}^{b}}\, G_{0\,+ +}\, \frac{\delta\,M_{1}^{+ i}}{\delta A_{+\,z}^{c}} \,-\nonumber \\
&-&\,G_{0\,+ +}\,\frac{\delta^{2}\,M_{L}^{+ +}}{\delta A_{+\,x}^{a} \delta A_{-\,y}^{b}}\, G_{0\,+ +}\, \frac{\delta\,M_{1}^{+ +}}{\delta A_{+\,z}^{c}}\,-\,
\,G_{0\,i +}\,\frac{\delta^{2}\,M_{L}^{+ +}}{\delta A_{+\,x}^{a} \delta A_{-\,y}^{b}}\, G_{0\,+ j}\, \frac{\delta\,M_{1}^{j i}}{\delta A_{+\,z}^{c}}\,=\,\nonumber \\
&=&\,G_{0\,+ +}\,\frac{\delta^{2}\,M_{L}^{+ +}}{\delta A_{+\,x}^{a} \delta A_{-\,y}^{b}}\, G_{0\,+ i}\, \frac{\delta\,M_{1,\,i -}}{\delta A_{+\,z}^{c}} \,+\,
G_{0\,i +}\frac{\delta^{2}\,M_{L}^{+ +}}{\delta A_{+\,x}^{a} \delta A_{-\,y}^{b}}\,G_{0\,+ +} \frac{\delta\,M_{1,\,- i}}{\delta A_{+\,z}^{c}} -\,\nonumber \\
&-& G_{0\,+ +} \frac{\delta^{2}\,M_{L}^{+ +}}{\delta A_{+\,x}^{a} \delta A_{-\,y}^{b}} G_{0\,+ +} \frac{\delta\,M_{1,\,- -}}{\delta A_{+\,z}^{c}} -
G_{0\,i +} \frac{\delta^{2}\,M_{L}^{+ +}}{\delta A_{+\,x}^{a} \delta A_{-\,y}^{b}} G_{0\,+ j} \frac{\delta\,M_{1,\,j i}}{\delta A_{+\,z}^{c}}\,.
\eeqar
The only non-zero contribution comes from the last term in the r.h.s. of the expression, see \eq{Ef2}-\eq{Ef3}.
We have:
\beq \label{D23}
G_{0\,i +} \frac{\delta^{2}\,M_{L}^{+ +}}{\delta A_{+\,x}^{a} \delta A_{-\,y}^{b}} G_{0\,+ j} \frac{\delta\,M_{1,\,j i}}{\delta A_{+\,z}^{c}}\,=\,\frac{g}{N}\,
G_{0\, i +}^{t w}\,\frac{\delta\,\Le U_{1}^{a_1 d a_2}\Ra^{+}_{w p}}{\delta A_{+\,x}^{a}}\,\frac{\delta\,  \D_{i}^{2} A_{-\,w}^{d}}{\delta A_{-\,y}^{b}}\,
G_{0\, + j}^{p t}\,\frac{\delta\,M_{1,\,j i,\,t}^{a_2 a_1}}{\delta A_{+\,z}^{c}}\,,
\eeq
see \eq{C4} above. With the help of \eq{C6} and \eq{C16} we write:
\beqar \label{D24}
&\,&-2\,\imath\, K^{a\,b\,c}_{x\,y\,z\,1,\,2}= -\frac{2 g^3}{N}\,f_{a_2 c a_1}\,G_{0\, i +}^{t w}\,\Le U_{2}^{a_1 d a_2 a_3}\Ra^{+ +}_{w p w_1}\,
\frac{\delta\, v_{+\,w_1}^{a_3\,cl}}{\delta\,A_{+\,x}^{a}}\,\frac{\delta\,  \D_{i}^{2} A_{-\,w}^{d}}{\delta A_{-\,y}^{b}}\,
G_{0\, + i}^{p t}\,\Le \delta^{2}_{t_{\bot} z_{\bot}}\,\delta_{t^{+} z^{+}}\,\D_{-\,t}\Ra\,=\,\nonumber \\
&=&-\frac{2 g^3}{N}\,f_{a_2 c a_1} G_{0\, i +}^{t w} \Le U_{2}^{a_1 b a_2 a}\Ra^{+ +}_{w p w_1}
\Le  \delta^{2}_{w_{1\,\bot} x_{\bot}} \delta_{w_{1}^{+} x^{+}} \Ra \Le  \delta^{2}_{w_{\bot} y_{\bot}} \delta_{w^{-} y^{-}} \D^{2}_{\bot\,w}\Ra
G_{0\, + i}^{p t} \Le \delta^{2}_{t_{\bot} z_{\bot}} \delta_{t^{+} z^{+}} \D_{-\,t}\Ra
\eeqar
where
\beqar\label{D25}
\Le U_{2}^{a_1 b a_2 a} \Ra_{w p w_1}\,& = &\,
tr[\,f_{a_1}\,G^{+}_{w p}\,f_{a_2}\,G^{+}_{p w_1}\,f_{a}\,O_{w_1}\,f_{b}\,O^{T}_{w}]\,+\,
tr[\,f_{a_1}\,G^{+}_{w w_1}\,f_{a}\,G^{+}_{w_1 p}\,f_{a_2}\,O_{p}\,f_{b}\,O^{T}_{w}]\,+\,\nonumber \\
& + &\,
tr[\,f_{a}\,G^{+}_{w_1 w}\,f_{a_1}\,G^{+}_{w p}\,f_{a_2}\,O_{p}\,f_{b}\,O^{T}_{w_1}]\,+\,
tr[\,f_{a_2}\,G^{+}_{p w}\,f_{a_1}\,G^{+}_{w w_1}\,f_{a}\,O_{w_1}\,f_{b}\,O^{T}_{p}]\,+\,\nonumber \\
& + &
tr[\,f_{a}\,G^{+}_{w_1 p}\,f_{a_2}\,G^{+}_{p w}\,f_{a_1}\,O_{w}\,f_{b}\,O^{T}_{w_1}]\,+\,
tr[\,f_{a_2}\,G^{+}_{p w_1}\,f_{a}\,G^{+}_{w_1 w}\,f_{a_1}\,O_{w}\,f_{b}\,O^{T}_{p}]\,.
\eeqar
Now we use the following identities for the requested traces:
\beq\label{D26}
f_{a_2 c a_1}\,tr[f_{a_1} f_{a_2} f_a f_b]\,=\,-\,\frac{N^2}{4}\,f_{a b c}
\eeq
see \eq{D6} above. Correspondingly, taking into account that
\beq\label{D261}
tr[f_{a_1} f_{a} f_{a_2} f_b]\,=\,\delta_{a_1 a}\delta_{a_2 b} \,+\,\delta_{a_1 b}\delta_{a a_2} \,+\,\frac{N}{4}\,\Le
 d_{a_1 a e} d_{a_2 b e}\,+\,d_{a_2 a e} d_{a_1 b e}\,-\,d_{a_1 a_2 e} d_{a b e}\, \Ra\,
\eeq
is symmetrical in respect to $a_1$ and $a_2$, we obtain
\beq\label{D2611}
f_{a_2 c a_1}\,tr[f_{a_1} f_{a} f_{a_2} f_b]\,=\,0\,.
\eeq
Therefore we have:
\beq\label{D27}
f_{a_2 c a_1}\,\Le U_{2}^{a_1 b a_2 a} \Ra_{w p w_1}\,=\,\frac{N^2}{4}\,f_{a b c}\,\Le\, G^{+\,0}_{w_1 w}\,G^{+\,0}_{w p}\,+\,G^{+\,0}_{p w}\,G^{+\,0}_{w w_1}\,-\,
G^{+\,0}_{w p}\,G^{+\,0}_{p w_1}\,-\,G^{+\,0}_{w_1 p}\,G^{+\,0}_{p w}\,\Ra\,.
\eeq
Let's again consider different contributions in \eq{D24} expression.
\begin{enumerate}
\item

The first term consists of $\theta^{+}_{w_1 w}\,\theta^{+}_{w p}\,$ theta functions. After the integration on $+$ components of coordinates we obtain the following integrals
with respect to $+$ components of momenta: 
\beq\label{D28}
\int\,d k_{1\,+}\int\,d k_{+} \frac{e^{-\imath(k_{+} - k_{1\,+})(x^{+} - z^{+})}}{\Le k_{+} + \imath\varepsilon \Ra\,\Le k_{+} - k_{1\,+} + \imath\varepsilon  \Ra\,
\Le k_{+} - k_{\bot}^{2}/2k_- + \imath\varepsilon \Ra\,\Le k_{1\,+} - k_{1\,\bot}^{2}/2k_- + \imath\varepsilon \Ra}\,,
\eeq
where the integration on $\delta(k_{-} - k_{1\,-})$ functions was performed.
First of all, we perform integration on $k_{1\,+}$ variable obtaining:
\beqar\label{D29}
&\,&-\,\int\,d k_{1\,+}\,\frac{e^{\imath\, k_{1\,+}(x^{+} - z^{+})}}{\Le k_{1\,+} - k_{+} - \imath\varepsilon \Ra\,\Le k_{1\,+} - k_{1\,\bot}^{2}/2k_- + \imath\varepsilon \Ra}\,=\,\nonumber \\
&=&\,-2\,\pi\,\imath\,\Le  \frac{\theta(x^{+} -z^{+})\,e^{\imath\, k_{+}(x^{+} - z^{+})}}{\Le k_{+} - k_{1\,\bot}^{2}/2k_- + \imath\varepsilon \Ra}\,+\,
\frac{\theta(z^{+} -x^{+})\,e^{ -\imath\, k_{1\,\bot}^{2}(z^{+} - x^{+})/2k_-}}{\Le k_{+} - k_{1\,\bot}^{2}/2k_- + \imath\varepsilon \Ra}\Ra\,.
\eeqar
The integration on $k_{+}$ variable in the second term of \eq{D29} provides $\theta(x^{+} -z^{+})$ function in the final answer, therefore this contribution is zero. 
The remaining integral is the following one:
\beqar\label{D30}
&\,&\int\,d k_{+}\,\frac{1}{\Le k_{+} + \imath\varepsilon \Ra\,\Le k_{+} - k_{\bot}^{2}/2k_- + \imath\varepsilon \Ra\,
\Le k_{+} - k_{1\,\bot}^{2}/2k_- + \imath\varepsilon \Ra}\,\propto\,\nonumber \\
&\propto&\,\frac{1}{k_{\bot}^2\,k_{1\,\bot}^{2}}\,+\,\frac{1}{k_{\bot}^{2}\,\Le k_{\bot}^2\,-\,k_{1\,\bot}^{2} \Ra }\,+\,\frac{1}{k_{1\,\bot}^{2}\,\Le k_{1\bot}^2\,-\,k_{\bot}^{2} \Ra }\,=\,0\,.
\eeqar

\item
The expression for the third term in \eq{D27} can be obtained from the previous one by the $k_{\bot},x\,\rightleftarrows\,k_{1\,\bot},z$ substitution, i.e. it is equal to zero as well.

\item
The next term we consider is proportional to $\theta^{+}_{p w}\,\theta^{+}_{w w_1}\,$ theta functions. This contributions is proportional to the following integral:
\beq\label{D32}
-\,\int\,d k_{1\,+}\int\,d k_{+} \frac{e^{-\imath(k_{+} - k_{1\,+})(x^{+} - z^{+})}}{\Le k_{+} - \imath\varepsilon \Ra\,\Le k_{+} - k_{1\,+} - \imath\varepsilon  \Ra\,
\Le k_{+} - k_{\bot}^{2}/2k_- + \imath\varepsilon \Ra\,\Le k_{1\,+} - k_{1\,\bot}^{2}/2k_- + \imath\varepsilon \Ra}\,
\eeq
or after the variables change to
\beq\label{D321}
\int\,d s_{+}\int\,d k_{+} \frac{e^{\imath\,s_{+}\,(x^{+} - z^{+})}}{\Le k_{+} - \imath\varepsilon \Ra\,\Le s_{+} + \imath\varepsilon  \Ra\,
\Le k_{+} - k_{\bot}^{2}/2k_- + \imath\varepsilon \Ra\,\Le k_{+} + s_{+} - k_{1\,\bot}^{2}/2k_- + \imath\varepsilon \Ra}\,.
\eeq
An integration on $k_{+}$ and subsequent integration with rspect to $s_{+}$ provides:
\beqar\label{D322}
&-&\,2\,\pi\,\imath\,\frac{2\,k_{-}}{k_{\bot}^{2}}\int\,d s_{+}\frac{e^{\imath\,s_{+}\,(x^{+} - z^{+})}}{\Le s_{+} + \imath\varepsilon  \Ra\,
\Le s_{+} - k_{1\,\bot}^{2}/2k_- + \imath\varepsilon \Ra}\,=\,\,\nonumber \\
&=&\,-\,(2\,\pi\,\imath\,)^{2}\,\frac{4\,k_{-}^2}{k_{\bot}^{2}\,k_{1\,\bot}^{2}}
\,\Le 1\,-\,e^{\imath\,\frac{k_{1\,\bot}^{2}}{2k_-}\,(x^{+} - z^{+})}\,\Ra\,\theta(z^{+}\,-\,x^{+})\,.
\eeqar
\item
The last term in \eq{D27} can be obtained from the \eq{D321} answer by changinh the overall sign of the expression and $k_{\bot},x\,\rightleftarrows\,k_{1\,\bot},z$ substitution.
\end{enumerate}
Finally, taking second and third term's contributions of \eq{D2} together, we obtain:
\beqar \label{D36}
&\,&-2\,\imath\, K^{a\,b\,c}_{x\,y\,z\,1,\,2}=\,\nonumber \\
&=& 
\frac{\imath\,g^3\,N}{(2\pi)^{5}}\,f_{a b c}\,\D_{i\,y}^{2}\,\Le\,
\theta(z^{+}\,-\,x^{+})\,\delta^{2}_{x_{\bot}\,y_{\bot}}\,
\int\, \frac{d k_{-}}{k_{-}}\,\int d^2 k_{\bot} \,\int d^{2} k_{1\,\bot}\,e^{-\imath\,(x^{i}\,-\,z^{i})\,(k_{i}\,-\,k_{1\,i})}\,
\frac{k_{i}\,k_{1\,i}}{(k_{i})^{2}\,(k_{1\,i})^{2}}\,\right.
\nonumber \\
&&\left.
\Le\,1\,-\,\frac{1}{2}\,e^{\imath\,\frac{k_{\bot}^{2}}{2k_-}\,(x^{+} - z^{+})}\,-\,\frac{1}{2}\,e^{\imath\,\frac{k_{1\,\bot}^{2}}{2k_-}\,(x^{+} - z^{+})}\,\Ra\,-\,
\theta(x^{+}\,-\,z^{+})\,\delta^{2}_{z_{\bot}\,y_{\bot}}\,
\int\, \frac{d k_{-}}{k_{-}}\,\int d^2 k_{\bot} \,\right.
\nonumber \\
&\,&\left.\,\int d^{2} k_{1\,\bot}\, e^{-\imath\,(z^{i}\,-\,x^{i})\,(k_{i}\,-\,k_{1\,i})}\,
\frac{k_{i}\,k_{1\,i}}{(k_{i})^{2}\,(k_{1\,i})^{2}}\,
\Le\,1\,-\,\frac{1}{2}\,e^{\imath\,\frac{k_{\bot}^{2}}{2k_-}\,(z^{+} - x^{+})}\,-\,\frac{1}{2}\,e^{\imath\,\frac{k_{1\,\bot}^{2}}{2k_-}\,(z^{+} - x^{+})}\,\Ra\,
\Ra\,.
\eeqar
Considering the integral with respect to $k_-$ variable only, we make the change of the variables cand expanding  the exponentials we obtain teh following the following integral:
\beq\label{D37}
\sum_{n=0}^{\infty}\,\int_{-\infty}^{\infty}\, dk_{-} \,\Le\,\frac{-\,\imath^{n+1}}{(n+1)!\,k_{-}^{n+2}}\,\Ra\,=\,0\,,
\eeq
therefore the whole \eq{D36} contribution is zero as well.

\subsection*{Third contribution}

 The only contribution of the last two terms in \eq{D2} can be written as:
\beq \label{D371}
-2\,\imath\, K^{a\,b\,c}_{x\,y\,z\,1,\,3}= 
G_{0\,k +}\,\frac{\delta M_{L}^{+ +}}{\delta A_{-\,y}^{b}}\, G_{0\,+ i}\, \frac{\delta M_{1}^{i j}}{\delta A_{+\,x}^{a}} \,
G_{0\,j p}\, \frac{\delta M_{1}^{p k}}{\delta A_{+\,z}^{c}} +
G_{0\,k +}\,\frac{\delta M_{L}^{+ +}}{\delta A_{-\,y}^{b}}\, G_{0\,+ i}\, \frac{\delta M_{1}^{i j}}{\delta A_{+\,z}^{c}} \,.
G_{0\,j p}\, \frac{\delta M_{1}^{p k}}{\delta A_{+\,x}^{a}} \,.
\eeq
The formulas we use here are \eq{C14}-\eq{C16} only, taking into account that only $G^{+\,0}_{w t}$ from \eq{C15} contributes in the final answer for the positive values of $k_-$, we obtain for the
momentum integrals in the first term of \eq{D37} expression:
\beq\label{D38} 
-\,\delta_{u^{+}\,x^{+}}\,\delta_{z^{+}\,s^{+}}\,\int\,\frac{d^4 k}{(2 \pi)^4}\,\frac{d^{4} k_{1}}{(2 \pi)^4}\,\frac{k_{i}\,k_{1\,i}}{k^{2}\,k_{1}^{2}\,k_{1\,-}}\,\theta(w^{+}\,-\,t^{+})\,
e^{-\,\imath\,k\,(s\,-\,t)\,-\,\imath\,k_1\,(w\,-\,u)}\,
\int\,\frac{d^4 k_2}{(2 \pi)^4}\,\frac{k_{2\,-}}{k_{2}^{2}}\,e^{-\,\imath\,k_2\,(u\,-\,s)}\,,
\eeq
the overall $-$ sign here is from the sign of $G^{+\,0}_{w t}$ Green's function.
The integration on the following coordinate variables  provides in turn:
\beq\label{D39}
\int\,d t^{+}\,\int\,d w^{+}\,\int d u^{-}\,\int d s ^{-}\,\Le \cdots \,\Ra\,=\,-\,\frac{(2\,\pi)^{3}\,\imath}{k_{+}\,-\,\imath\,\varepsilon}\,\delta_{k_{+} k_{1\,+}}\,
\delta_{k_{2\,-} k_{1\,-}}\,\delta_{k_{-} k_{2\,-}}\,\,.
\eeq
Therefore, an integration in respect to $k_{-}\,k_{+}\,k_{1\,-}\,k_{1\,+}\,$ gives:
\beqar\label{D40}
&\,&\frac{k_{i} k_{1\,i}}{(2 \pi)^{8}}\,\int\,\frac{d k_{-}}{k_{-}}\,\int \frac{d k^{+}}{(2 k_-)^2}\,\frac{e^{-\,\imath\,k_+\,(z^{+}\,-\,x^{+})}}
{\Le k_{+} - \imath\varepsilon \Ra\,\Le k_{+} - k_{\bot}^{2}/2k_- + \imath\varepsilon \Ra\,\Le k_{+} - k_{1\,\bot}^{2}/2k_- + \imath\varepsilon \Ra}\,=\,\nonumber \\
&=&\,\frac{2 \pi \imath}{(2 \pi)^{8}}\,\frac{k_{i} k_{1\,i}}{k_{\bot}^{2}\,k_{1\,\bot}^{2}}\,\int\,\frac{d k_{-}}{k_{-}}\,\theta(x^{+} - z^{+})\,.
\eeqar
Correspondingly, the integral on $k_{2}$ variable provides:
\beq\label{D4111}
\int \frac{d k_{2\,+}}{2 \,(2 \pi)^{4}}\,
\frac{e^{-\,\imath\,k_{2\,+}\,(x^{+}\,-\,z^{+})}}{k_{2\,+} - k_{2\,\bot}^{2}/2k_- + \imath\varepsilon }\,=\,
\frac{-\,\pi\,\imath}{(2 \pi)^{4}}\,\theta(x^{+} - z^{+})\,e^{-\imath\frac{k_{2\,\bot}^{2}}{2 k_{-}}\,(x^{+}\,-\,z^{+})}\,.
\eeq
Now, taking all two terms of \eq{D371} together, we obtain:
\beqar \label{D42}
&\,& -2\,\imath\, K^{a\,b\,c}_{x\,y\,z\,1,\,3}\,=
\,\frac{\imath\,g^3\,N}{2 (2 \pi)^{7}}\,f_{a b c}\,
\D_{i\,y}^{2}\,\Le\,
\theta(z^{+} - x^{+})\,\int \frac{d k_{-}}{k_{-}}\,\int d^{2} k\bot\,\int d^{2} k_{1\,\bot}\,\int d^{2} k_{2\,\bot}\,\frac{k_{i} k_{1\,i}}{k_{\bot}^{2}\,k_{1\,\bot}^{2}}\,
\right.\nonumber \\
&\,& \left.
e^{-\,\imath\,k_i\,(x^{i}\,-\,y^{i})-\,\imath\,k_{1\,i}\,(y^{i}\,-\,z^{i})
-\,\imath\,k_{2\,i}\,(z^{i}\,-\,x^{i})\,-\,\imath\frac{k_{2\,\bot}^{2}}{2 k_{-}}\,(z^{+}\,-\,x^{+})}\,-\,\theta(x^{+} - z^{+})\,\int \frac{d k_{-}}{k_{-}}\,\int d^{2} k_{\bot}\,
\right. \nonumber \\
&\,&\,\left.
 \int d^{2} k_{1\,\bot}\,\int d^{2} k_{2\,\bot}\,\frac{k_{i} k_{1\,i}}{k_{\bot}^{2}\,k_{1\,\bot}^{2}}\,
e^{-\,\imath\,k_i\,(z^{i}\,-\,y^{i})-\,\imath\,k_{1\,i}\,(y^{i}\,-\,x^{i})
-\,\imath\,k_{2\,i}\,(x^{i}\,-\,z^{i})\,-\,\imath\frac{k_{2\,\bot}^{2}}{2 k_{-}}\,(x^{+}\,-\,z^{+})}\,
\Ra\,.
\eeqar
We note, that for the case when $x^{+}\,=\,z^{+}$ we obtain $K^{a\,b\,c}_{x\,y\,z\,1,\,3}\,=\,0\,$ as it must be. 
We also note, the first term
in \eq{C15} will contribute to the final answer as well. Therefore, 
the whole \eq{D42} contribution must be doubled and we will obtain for it:
\beqar \label{D43}
&\,& -2\,\imath\, K^{a\,b\,c}_{x\,y\,z\,1,\,3}\,=\, \frac{\imath\,g^3\,N}{(2 \pi)^{5}}\,f_{a b c}\,\delta^{2}_{x_{\bot}\,z_{\bot}}\,
\D_{i\,y}^{2}\,\\
&\,&\Le\,
\theta(z^{+} - x^{+})\,\int \frac{d k_{-}}{k_{-}}\,
\int d^{2} k_{\bot}\,\int d^{2} k_{1\,\bot}\,\frac{k_{i} k_{1\,i}}{k_{\bot}^{2}\,k_{1\,\bot}^{2}}\,
e^{-\,\imath\,k_i\,(x^{i}\,-\,y^{i})-\,\imath\,k_{1\,i}\,(y^{i}\,-\,z^{i})}\,-\, \right. \nonumber \\
&-&\,\left.
\theta(x^{+} - z^{+})\,\int \frac{d k_{-}}{k_{-}}\,\int d^{2} k_{\bot}\,
\int d^{2} k_{1\,\bot}\,\frac{k_{i} k_{1\,i}}{k_{\bot}^{2}\,k_{1\,\bot}^{2}}\,
e^{-\,\imath\,k_i\,(z^{i}\,-\,y^{i})-\,\imath\,k_{1\,i}\,(y^{i}\,-\,x^{i})}\,,
\Ra\, \nonumber .
\eeqar  
where again the only first term of the expansion of $e^{-\frac{\imath}{k_-}}$ function remain in the final expression.

\end{document}